\documentclass[aps,prx,twocolumn,longbibliography]{revtex4-1}
\usepackage{bm}
\usepackage{color} 
\usepackage{graphicx}
\usepackage{epsfig}
\usepackage{hyperref}
\usepackage{natbib}
\usepackage{amsmath}
\usepackage{amssymb}
\usepackage{bbm}

\AtBeginDocument{%
    \newwrite\bibnotes
    \def\bibnotesext{Notes.bib}
    \immediate\openout\bibnotes=\jobname\bibnotesext
    \immediate\write\bibnotes{@CONTROL{REVTEX41Control}}
    \immediate\write\bibnotes{@CONTROL{%
    apsrev41Control,author="08",editor="1",pages="0",title="0",year="1"}}
     \if@filesw
     \immediate\write\@auxout{\string\citation{apsrev41Control}}%
    \fi
}%

\newcommand{\rmi}{{\rm i}}

\newcommand {\e}{{\rm e}}

\graphicspath{{./figs/},{./figsNL/}}

\begin{document}

\title{Nonlinear dynamical Casimir effect and Unruh entanglement\\ in waveguide QED with parametrically modulated coupling}

\author{Egor S. Vyatkin}
\affiliation{Ioffe Institute, St. Petersburg 194021, Russia}

\author{Alexander V. Poshakinskiy}
\affiliation{ICFO-Institut de Ciencies Fotoniques, The Barcelona Institute of Science and Technology, 08860 Castelldefels (Barcelona), Spain}
\email{Alexander.Poshakinskiy@icfo.eu}

\author{Alexander N. Poddubny}
\affiliation{Department of Physics of Complex Systems, Weizmann Institute of Science, Rehovot 7610001, Israel}
\email{poddubny@weizmann.ac.il}
\begin{abstract}

We study theoretically an array of two-level qubits moving relative to a one-dimensional waveguide. This motion can be implemented mechanically or simulated via the modulation of the couplings between the qubits and the waveguide. When the frequency of this motion approaches twice the qubit resonance frequency, it induces parametric generation of photons and excitation of the qubits.
The proposed quantum optomechanical system offers a plethora of possibilities for exploring various quantum electrodynamics phenomena. However, their theoretical analysis is challenging due to the presence of quantum nonlinearity, a continuum of propagating photonic modes, and the excitation of strongly nonequilibrium qubit states, which make many conventional analytical tools inapplicable. To address these challenges, we develop a comprehensive general theoretical framework that incorporates both perturbative diagrammatic techniques and a rigorous master-equation approach.
Our calculations reveal several intriguing effects, including the directional dynamical Casimir effect, where momenta of emitted photon pairs are correlated, and the waveguide-mediated collective Unruh effect, where motion drives the qubits to a nontrivial steady state that can be entangled and exhibit phase transitions. Additionally, we examine the radiation back-action on the qubit motion, which becomes particularly pronounced when subradiant modes in the qubit array are excited. The back-action can significantly alter the mechanical spectra, potentially leading to the formation of hybrid phonon-biphoton modes.

\end{abstract}

\date{\today}

\maketitle

\section{Introduction}\label{sec:intro}

Quantum electrodynamics (QED) predicts several intriguing effects that have not been directly observed due to their extreme weakness. One such effect is the dynamical Casimir effect (DCE), which posits that a mirror undergoing oscillatory motion in a vacuum emits light~\cite{Moore1970,DeWitt1975}. Consequently, it experiences a radiation friction force~\cite{Fulling1976,Braginsky1991}, which is the quantum counterpart of the radiation recoil experienced by charged particles in classical electrodynamics. DCE emission and friction are predicted for almost any motion of neutral polarizable objects, including oscillating semitransparent mirrors~\cite{Lambrecht1996}, sliding parallel plates~\cite{Teodorovich1978,Levitov1989}, and rotating bodies~\cite{Zeldovich1970,Maghrebi2011}. Recently, the backaction of DCE forces on quantized mechanical motion in optomechanical systems has been investigated~\cite{Macri2018,DiStefano2019,Butera2019,DelGrosso2019}. However, all DCE phenomena, being both quantum and relativistic in nature, remain exceedingly weak in realistic experimental setups, where achieving high mechanical frequencies and relativistic velocities is a challenge~\cite{Dodonov2020review}.

The Unruh effect is a related quantum relativistic phenomenon that predicts that an observer accelerating uniformly in a vacuum perceives an equivalent of a thermal bath with a finite temperature, termed the Unruh temperature~\cite{Unruh1976,Crispino2008,BenBenjamin2019}. As a result, an accelerating two-level detector has a finite probability of being found in its excited state~\cite{Unruh1976,DeWitt1979}, with excitation process accompanied by photon emission, in a manner reminiscent of the anomalous Doppler effect~\cite{Ginzburg1987}. This emission can be interpreted as a consequence of the DCE. The Unruh effect is enhanced when acceleration occurs inside a cavity~\cite{Scully2003,Belyanin2006,Stargen2022}. The effect was also studied for extended systems~\cite{Lima2019}, where its effect on entanglement was considered~\cite{Tian2012,Mukherjee2023}. The concept has further been extended to motions with time-dependent acceleration~\cite{Obadia2007,Kothawala2010,Barbado2012}, including oscillatory motion~\cite{Doukas2013,Lin2017}. Oscillatory motion within a cavity for a single two-level system~\cite{Agusti2021,Wang2021} and ensembles of such systems have also been studied~\cite{Dodonov2018,Dodonov2022}.

Over the past decade, microwave waveguides coupled to superconducting quantum interference devices (SQUIDs) and transmon qubits have become powerful tools for simulating cavity quantum electrodynamics~\cite{Blais2004,Blais2021} and quantum electrodynamics in 1+1 dimensions~\cite{Astafiev2010,Nation2012}. In these setups, SQUIDs with externally controlled magnetic flux can be employed to simulate motion. Notably, in the case of simulated motion, the distinction between the DCE and the Unruh effect becomes blurred, as both phenomena are essentially forms of parametric excitation~\cite{Blais2021}. We will refer to photon emission as the DCE, and to changes in the emitter state as the Unruh effect.

To simulate a moving mirror, it was proposed to place a modulated SQUID at the end of a waveguide, which results in dynamically changing boundary conditions~\cite{Johansson2009,Nori2010,Nori2018}. This led to the experimental observation of DCE-like radiation~\cite{Wilson2011}. The strongly non-classical nature of DCE emission has also been demonstrated~\cite{Johansson2013,Sabin2015}, with potential applications in inducing qubit entanglement~\cite{Sun2024}.

Simulating the motion of a two-level detector in the Unruh effect requires dynamical tuning of the coupling between a qubit and a waveguide or resonator. This can be achieved by using a SQUID-controlled coupler~\cite{Allman2010,Bialczak2011}, or by employing a pair of interacting transmons as a single tunable qubit, which allows for independent control over the qubit resonance frequency and its coupling to the resonator through two separate magnetic fluxes~\cite{Gambetta2011,Srinivasan2011}. These techniques have been also proposed to simulate qubit motion inside a microwave resonator, which leads to DCE emission, excitation of the qubit~\cite{Felicetti2015,Sabin2017}, and qubit entanglement~\cite{Rossatto2016,GarciAlvarez2017,Agusti2019,AbdelKhalek2019}.

Flux-tunable transmons enable temporal modulation of the qubit resonance frequency~\cite{DiCarlo2009,Rol2019}, allowing for directional light scattering~\cite{Redchenko2022}, the generation of entangled frequency combs~\cite{Ilin2023}, and DCE-like parametric photon generation~\cite{Vyatkin2023}. Parametric modulation in superconducting qubit systems has also been utilized to realize quantum gates~\cite{McKay2016,Roth2017} and achieve chiral behavior~\cite{Joshi2023}.


In this paper, we investigate an array of two-level qubits undergoing oscillatory motion and interacting with photons in a one-dimensional waveguide. The system possesses unique features that render previously developed theoretical techniques inapplicable. (i) The motion of a qubit relative to the waveguide leads to parametric coupling with a continuum of photonic modes, making it impossible to calculate the full system evolution directly, as is commonly done in systems with parametric coupling to a single-mode resonator~\cite{DeLiberato2009,Vacanti2012,Bastidas2012,Chitra2015,Groszkowski2020}. (ii) The system is strongly nonlinear at the two-photon level, as a qubit can scatter one photon but not a pair of photons simultaneously~\cite{Yudson1984,Shen2007}. Given that photons in the DCE are produced in pairs, this nonlinearity must be considered even in regimes of weak motion. Therefore, field quantization~\cite{Moore1970,Dodonov2020review} cannot be performed explicitly, nor can be applied scattering approaches~\cite{Lambrecht1996,Maghrebi2013,Kerker2019}, which were developed for macroscopic structures and assume linear optical response. (iii) Furthermore, under strong resonant motion, the qubit array may enter highly excited states, necessitating the use of non-equilibrium many-body theoretical approaches~\cite{Kamenev2011}.

The aim of this paper is to develop theoretical frameworks that are suitable for such open, intrinsically nonlinear systems under strongly non-equilibrium conditions. We demonstrate that the unique features of this setup lead to novel manifestations of the DCE and Unruh effects. In particular, (i) parametric long-range coupling via the waveguide~\cite{Karnieli2024} enables tunable directional or direction-entangled DCE emission, (ii) excitation of correlated many-body states, including subradiant states~\cite{Molmer2019,Ke2019,Poshakinskiy2021dimer}, gives rise to strong optomechanical backaction forces, and (iii) proper engineering of strong parametric coupling allows the realization of dissipative phase transitions in many-qubit arrays~\cite{Kessler2012,Munos2019}.

The remainder of the paper is organized as follows. In Sec.~\ref{sec:Model}, we present the theoretical model that describes an array of atoms moving relative to a waveguide. Next, we explore the dynamics of the system using two different approaches.
First, in Sec.~\ref{sec:DCEperp}, we develop a perturbative diagrammatic approach. This method is applicable in the regime of weak modulation and provides relatively simple expressions for DCE emission from arbitrary qubit arrays. Additionally, we use this approach to describe how subradiant two-photon states can enhance DCE emission.
Second, in Sec.~\ref{sec:Unruh}, we develop an approach based on the master equation. This method is valid even under strong modulation and accounts not only for DCE emission but also for changes in the qubit states, corresponding to the Unruh effect. However, due to its computational complexity, this approach is limited to arrays with a small number of qubits. We use this method to investigate motion-induced qubit entanglement and squeezing, emission directivity, and correlations between emitted photons.
Then, in Sec.~\ref{sec:Back}, we study the backaction of DCE radiation on mechanical motion. We demonstrate the formation of hybrid modes, which are mixtures of two photons and one phonon, and analyze how these modes influence photon statistics.
Finally, we summarize our findings in Sec.~\ref{sec:Sum}.

\begin{figure}[tb]
 \centering\includegraphics[width=.99\columnwidth]{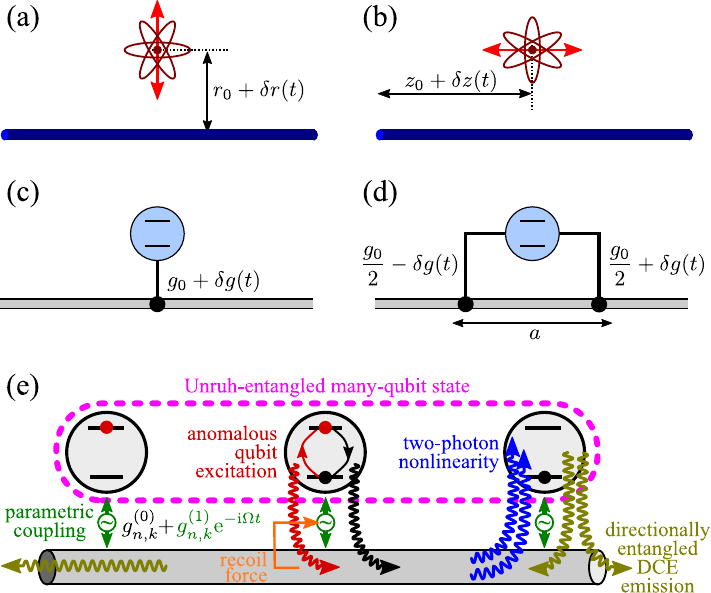}
 \caption{(a),(b) An elementary unit of the proposed setup: an atom moving periodically with respect to a waveguide, giving rise to the parametric coupling. The motion can happen in the direction perpendicular (a) or parallel (b) to the waveguide. (c),(d) Emulation  of mechanical motion in a superconducting circuit with transmons, where the couplings between the transmon and the waveguide are modulated in time. (e) The considered setup -- a qubit array parametrically coupled to a waveguide -- and the phenomena that it can realize.}\label{fig:emumov}
\end{figure}
\section{Model}\label{sec:Model}

We consider a waveguide QED setup where the coupling between atoms and waveguide photons, denoted by $g_k(t)$, is modulated over time at the parametric resonance frequency $\Omega \approx 2\omega_0$, where $\omega_0$ is the atomic resonance frequency. For real atoms trapped near a fiber, this modulation could be achieved through the periodic mechanical motion of the atoms relative to the waveguide. The motion can occur either perpendicular or parallel to the waveguide, as shown in Figs.~\ref{fig:emumov}(a) and (b), leading to:
\begin{align}
g_k(t) = g_0[r(t)] \e^{\rmi k z(t)} \,,
\end{align}
where $r(t)$ is the distance between the waveguide and the atom, $z(t)$ is the position of the atom along the waveguide, and $k$ is the photon wave vector. The modulation of $r(t)$ affects the amplitude of the coupling parameter $g_k(t)$, while $z(t)$ influences its phase. A key distinction between these two types of modulation is that the induced variation in the coupling parameter $g_k(t)$ is $k$-even in the former case and $k$-odd in the latter. This difference leads to distinct emission patterns and determines which atomic states can be excited.
{
For both types of motion, the efficiency of parametric excitation is determined by the relative amplitude of the modulation $v \equiv \delta g/g \sim ku \sim n_{\rm eff}\dot{u}/2c$, where $u$ and $\dot{u}$ represent the motion amplitude and velocity, respectively, and $n_{\rm eff}$ is the effective refractive index of the waveguide mode. Since $n_{\rm eff}$ can be quite high, large values of $v$ can potentially be achieved before reaching the limit $\dot{u} < c$. Although such rapid oscillatory motion will likely induce parametric modulation of the internal atomic structure as well, we will not consider such effects here, instead focusing on the modulation of the coupling.
}

Implementing parametric modulation through mechanical motion at optical frequencies is challenging. However, radio-frequency setups using artificial atoms --- such as transmon qubits --- offer much greater flexibility~\cite{Allman2010,Bialczak2011,Gambetta2011,Srinivasan2011,Kotler_2021,Blais2021}. In particular, modulating the coupling between a transmon qubit and the microwave waveguide, as illustrated in Fig.~\ref{fig:emumov}(c), emulates atomic motion in the radial direction. A giant artificial atom~\cite{Kockum2020,Du2022}, shown in Fig.~\ref{fig:emumov}(d), can emulate the motion of a real atom along the waveguide. Specifically, if the two links are modulated as $g_{1,2} = g_0/2 \mp \delta g(t)$, where $a$ is the distance between the points where the links connect to the waveguide, the total coupling is given by
$
g_k = g_1 \e^{-\rmi k a/2} + g_2 \e^{\rmi k a/2}
$.
For small $a \ll 1/k$, we have 
\begin{align} \label{eq:smalla}
g_k \approx g_0 + \rmi k a\, \delta g(t) \approx  g_0 \e^{\rmi k u(t)}\,,
\end{align}
where $u(t) = (a/g_0) \delta g(t)$ emulates the mechanical motion along the waveguide.

Regardless of the type of modulation or its physical realization, an array of qubits coupled to a waveguide is described by the Hamiltonian~\cite{Roy2017,Chang2018,Sheremet2023}
\begin{align}
H &= \sum_k \omega_k a_k^\dag a_k + \sum_n \omega_0 b_n^\dag b_n + V \,,\\ \nonumber
V&= \sum_{k,n} [b_n + b_n^\dag]\left[ g_{n,k}^*(t)\, a_k \e^{\rmi k z_n} + g_{n,k}(t)\, a_k^\dag \e^{-\rmi k z_n}\right] \,,
\end{align}
where $\hbar = 1$ is assumed, $\omega_k = c |k|$ represents the photon dispersion, $\omega_0$ is the qubit transition frequency, $a_k$ is the photon annihilation operator, $b_n$ is the lowering operator of the $n$-th qubit, and the time dependence of the coupling parameters $g_{n,k}(t)$ is explicitly introduced. Assuming this dependence is periodic, we decompose it into a Fourier series
\begin{align}
g_{n,k}(t) = \sum_p g^{(p)}_{n,k} \e^{-\rmi p \Omega t} \,.
\end{align}
We assume that the parametric resonance condition  $\Omega \approx 2\omega_0$ is fulfilled and apply the rotating-wave approximation (RWA). Retaining only the resonant terms, which oscillate at frequencies $|\Omega - \omega_0 - \omega_k| \ll \omega_0$, we obtain
\begin{align}\label{eq:Vrwa}
V_{\rm RWA} = \sum_{k,n} \left[  (g_{n,k}^{(0)*} b_n^\dag+ g_{n,k}^{(1)*} b_n \e^{\rmi\Omega t})  a_k \e^{\rmi k z_n} +{\rm h.c.}  \right] \,,
\end{align}
where all other harmonics $g^{(p)}_{n,k}$, except those with $p = 1$, are found to be irrelevant. Note that the explicit time dependence can be entirely eliminated from the Hamiltonian by transitioning to a frame rotating with the frequency $\Omega/2$.

 \section{DCE emission: perturbative approach}\label{sec:DCEperp}

The interaction Hamiltonian Eq.~\eqref{eq:Vrwa} can be separated into time-independent and time-dependent parts, $V_{\rm RWA} = V_0 + V_1 \e^{-\rmi \Omega t} + V_1^\dag \e^{\rmi \Omega t}$, where
\begin{align}
&V_0 = \sum_{k,n} \left[ g_{n,k}^{(0)*} b_n^\dag a_k \e^{\rmi k z_n} + g_{n,k}^{(0)} b_n a_k^\dag \e^{-\rmi k z_n}\right], \label{eq:V0}\\
&V_1 = \sum_{k,n} g_{n,k}^{(1)} b_n^\dag a_k^\dag \e^{-\rmi k z_n}. \label{eq:V1}
\end{align}
Here, $V_0$ has the usual form of interaction in waveguide QED~\cite{Roy2017,Chang2018,Sheremet2023}, which describes the conversion of a photon into a qubit excitation and vice versa. 
The term $V_1$ describes an anomalous qubit excitation process accompanied by photon emission rather than photon absorption, as illustrated by the red arrows in Fig.~\ref{fig:emumov}(e). This process is conceptually similar to the anomalous Doppler effect observed for atoms moving with superluminal velocity~\cite{Ginzburg1987}.

We will assume that $g^{(0)}_{k,n} = g_0$ is the same for all qubits and does not depend on the direction of $k$, i.e., in the absence of modulation, the system is not chiral. In this section, we also assume a modulation corresponding to motion along the waveguide,
\begin{align}\label{eq:g1mo}
g^{(1)}_{k,n} = \rmi k u_n g_0 \,,
\end{align}
where $u_n$ is the motion amplitude of the $n$-th atom, and $k|u_n| \ll 1$. We will show that this type of motion enables the excitation of dark double-excited states, while motion perpendicular to the waveguide does not. 

To describe the emergent DCE emission, we treat the parametric generation term $V_1$ in Eq.~\eqref{eq:V1} as a perturbation, considering it only to the first order. We will develop a diagrammatic technique similar to that used to describe two-photon scattering in waveguide QED~\cite{Ke2019,Sheremet2023}; see Appendix~\ref{app:Dia} for details.

\subsection{Single qubit}\label{sec:Low1}

First, we apply a perturbative approach to describe the simplest problem: the DCE emission from a single qubit vibrating along the waveguide. A non-perturbative treatment of this problem is provided in Appendix~\ref{app:1mov}.

After a qubit excitation and a photon are generated by the parametric generation term $V_1$, Eq.~\eqref{eq:V1}, the qubit can emit a second photon, as shown by the red and black wavy arrows in Fig.~\ref{fig:emumov}(e). For a single qubit vibrating along the waveguide, this is the only process contributing to DCE emission; see Fig.~\ref{fig:dia}(a) in Appendix~\ref{app:Dia} for details. The corresponding two-photon amplitude reads
\begin{align}\label{eq:psi1}
\psi_{\sigma_1,\sigma_2}& (\omega_1,\omega_2) \\\nonumber
&=- k_0 u_1 \Gamma_0 \left( \frac{\sigma_2}{\omega_1-\omega_0+\rmi\Gamma_0}+\frac{\sigma_1}{\omega_2-\omega_0+\rmi\Gamma_0}\right) 
 \,,
\end{align}
where $\omega_{1,2}$ are the frequencies of the two photons, which satisfy $\omega_1 + \omega_2 = \Omega$, $\sigma_{1,2} = \pm 1$ denote the photon propagation directions, and we use the Markovian approximation for the photon wave vectors $k_{1,2} \approx \sigma_{1,2} k_0$, with $k_0 = \omega_0/c$. Equation~\eqref{eq:psi1} is consistent with the previous result for DCE emission by an oscillating point scatterer with arbitrary linear polarizability in a 1D setup~\cite{Lambrecht1996}. This indicates that for a single qubit vibrating along the waveguide with a small amplitude, its optical nonlinearity is, in fact, not manifested.

\begin{figure}
    \centering
    \includegraphics[width=0.9\columnwidth]{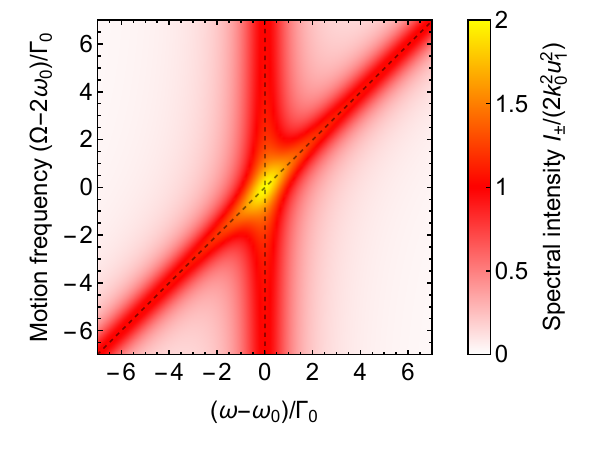}
    \caption{Color map showing the spectral density of the emission rate $I_\pm(\omega)$ for a single moving qubit, calculated from Eq.~\eqref{eq:I1}, as a function of $\omega$ and the mechanical motion frequency $\Omega$. Dashed lines indicate the positions of resonances for the two emitted photons.}
    \label{fig:I1}
\end{figure}

The spectrum of the photons emitted in the $\sigma$ direction is calculated as
\begin{align}\label{eq:I}
I_\sigma(\omega) = \left[ |\psi_{\sigma,\sigma}(\omega,\Omega-\omega)|^2 + |\psi_{\sigma,-\sigma}(\omega,\Omega-\omega)|^2 \right].
\end{align} 
Substituting Eq.~\eqref{eq:psi1} into Eq.~\eqref{eq:I}, we obtain
\begin{align}\label{eq:I1}
I_\sigma(\omega) &= 2|u_1|^2 k_0^2\Gamma_0^2 \\ \nonumber 
&\times \left[ \frac{1}{(\omega-\omega_0)^2+\Gamma_0^2} + \frac{1}{(\Omega-\omega-\omega_0)^2+\Gamma_0^2}\right].
\end{align}
The plot of $I_\sigma(\omega)$, which is the same for both directions, is shown in Fig.~\ref{fig:I1} as a function of the modulation frequency $\Omega$ and the emitted photon frequency $\omega$. The two peaks in the spectrum correspond to the cases when either the frequency of the detected photon $\omega$ or that of the complementary photon $\Omega-\omega$ matches the resonance frequency $\omega_0$. The two resonances do not interfere but simply add up. The resulting total photon emission rate $W  = W_+ + W_-$, where $W_\sigma = \int I_\sigma (\omega) d\omega/(2\pi)$, reads
\begin{align}\label{eq:W1}
W = 4|u_1|^2 k_0^2\Gamma_0.
\end{align}
Note that it does not have a resonant dependence on the modulation frequency $\Omega$.

The DCE emission has non-classical properties. Similar to other parametric generation processes, the emitted photons are squeezed~\cite{Scully1997book}. The squeezing of the field emitted in the direction $\sigma$ at the central frequency $\omega = \Omega/2$ can be described by the parameter $\xi$, which up to order $|u|^2$ reads
\begin{align}
\xi_{\sigma} = 1 - 2 | \psi_{\sigma,\sigma}(\tfrac{\Omega}{2},\tfrac{\Omega}{2})| + 2 I_{\sigma}(\tfrac{\Omega}{2}) + O(|u|^3),
\end{align}
cf. Eq.~\eqref{eq:K0} in Appendix~\ref{app:Emi}. Substituting Eqs.~\eqref{eq:psi1},~\eqref{eq:I1} into this expression and assuming $\Omega = 2\omega_0$, we obtain  
\begin{align}\label{eq:Xi1ser}
\xi_{\sigma}  &=  1 - 4 k_0 |u_1| + 8 k_0^2 |u_1|^2 + O(u_1^3).
\end{align} 
Note that in the regime of small motion amplitude $k_0 |u_1| \ll 1$, the squeezing is weak. Strong squeezing is achieved for large motion amplitudes $k_0 u_1 \sim 1$, which is beyond the applicability of the perturbative approach and is considered in Appendix~\ref{app:1mov}.

The photon statistics of the DCE emission are also non-classical. To quantify this, we perform the Fourier transform of Eq.~\eqref{eq:psi1}, yielding the two-photon wave function in real time:
\begin{align}\label{eq:psit}
&\psi_{\sigma_1,\sigma_2}(t_1,t_2) = \rmi \omega_0 u_1 \Gamma_0
\left[ \sigma_2 \theta(t_1-t_2) \e^{-[\rmi (\omega_0-\frac{\Omega}{2}) + \Gamma_0](t_1-t_2)} \right.
\nonumber\\ 
&\quad \left. + \sigma_1 \theta(t_2-t_1) \e^{-[\rmi (\omega_0-\frac{\Omega}{2}) + \Gamma_0](t_2-t_1)} \right] 
\e^{\rmi\frac{\Omega(t_1+t_2)}{2}}.
\end{align}
The (unnormalized) second-order correlation function of the emitted photons is given by $G^{(2)}_{\sigma_1,\sigma_2}(t_1-t_2) = |\psi_{\sigma_1,\sigma_2}(t_1,t_2)|^2$ and depends only on the delay time $t_1-t_2$. For the photons emitted in the same direction, $G^{(2)}_{\sigma,\sigma}(0) = (\omega_0 u_1 \Gamma_0)^2$ is finite. For the photons emitted in opposite directions, the wave function in Eq.~\eqref{eq:psit} has a discontinuity at $t_1-t_2 = 0$. This discontinuity originates from the rotating-wave approximation (RWA) and is, in fact, smoothed on a time scale of $\sim 1/\omega_0$, leading to $G^{(2)}_{\sigma,-\sigma}(0) = 0$, which can also be proven directly using the general result from Ref.~\cite{Lambrecht1996}. Thus, if the two photons are emitted within the same interval $\sim 1/\omega_0$, they propagate in the same direction, i.e., they are in the directionally entangled Bell state $|++\rangle + |--\rangle$.

\subsection{Qubit arrays}

In an array of vibrating qubits, the photon pairs emitted by one qubit can excite other qubits and interact due to the qubit nonlinearity, as indicated by the blue wavy arrows in Fig.~\ref{fig:emumov}(e). This interaction leads to a rather cumbersome expression for the two-photon amplitude, as shown in Eq.~\eqref{eq:psiN} in Appendix~\ref{app:Dia}. Here, we instead focus on the total emission rate $W$, which can be calculated either by integrating the emission spectrum or by applying the optical theorem:
\begin{align}\label{eq:WN}
W = - 4 \, \text{Im\,} \Sigma_{nm}(\Omega) u_n^* u_m \,,
\end{align}
where $\Sigma_{ij}(\Omega)$ is the self-energy of the mechanical motion. The calculation presented in Appendix~\ref{app:Dia} yields:
\begin{align}\label{eq:sigNeig}
\Sigma_{nm}(\Omega) =  k_0^2 \Gamma_0 \left( \Gamma_0\sum_{\nu} \frac{ \alpha^{(\nu)}_n  \alpha^{(\nu)}_m   }{\Omega - E^{(\nu)}} -\rmi \delta_{nm}  \right) \,.
\end{align}
Here, $\nu$ enumerates all double-excited states of the qubit array, which are described by complex energies $E^{(\nu)}$ and symmetric wave functions $\psi^{(\nu)}_{nm}$. The coefficients
\begin{align}\label{eq:alpha}
\alpha^{(\nu)}_n= \sum_{m} \psi^{(\nu)}_{nm} \ \text{sign}(z_n-z_m)\,\e^{\rmi k_0|z_n-z_m|} 
\end{align}
quantify the efficiency of mode excitation by the vibration of the $n$-th qubit along the waveguide. Note that $\sum_n \alpha^{(\nu)}_n = 0$. Therefore, if all the qubits vibrate uniformly, i.e., $u_i = u_1$, the total emission is equivalent to that from $N$ independent qubits, $W = N k_0^2 |u_1|^2 \Gamma_0$, and does not exhibit any resonant features.

Now, we apply the general results described above to equidistant arrays with two and four qubits. For non-trivial vibrational modes, resonances do arise in the emission rate dependence on the vibration frequency $\Omega$.

\subsubsection{Two qubits}

\begin{figure}
    \centering
    \includegraphics[width=0.9\columnwidth]{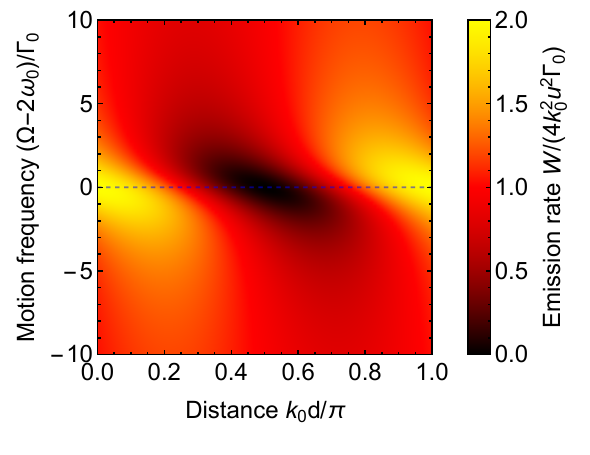}
    \caption{
    The total photon emission rate for a pair of qubits moving with opposite phases plotted after Eqs.~\eqref{eq:WN} and~\eqref{eq:q2sig} as a function of the motion frequency $\Omega$ and the distance between the qubits $d$. The dashed line represents the energy of the double-excited state of the two qubits.
    }
    \label{fig:2}
\end{figure}
\begin{figure*}
    \centering
    \includegraphics[width=0.9\textwidth]{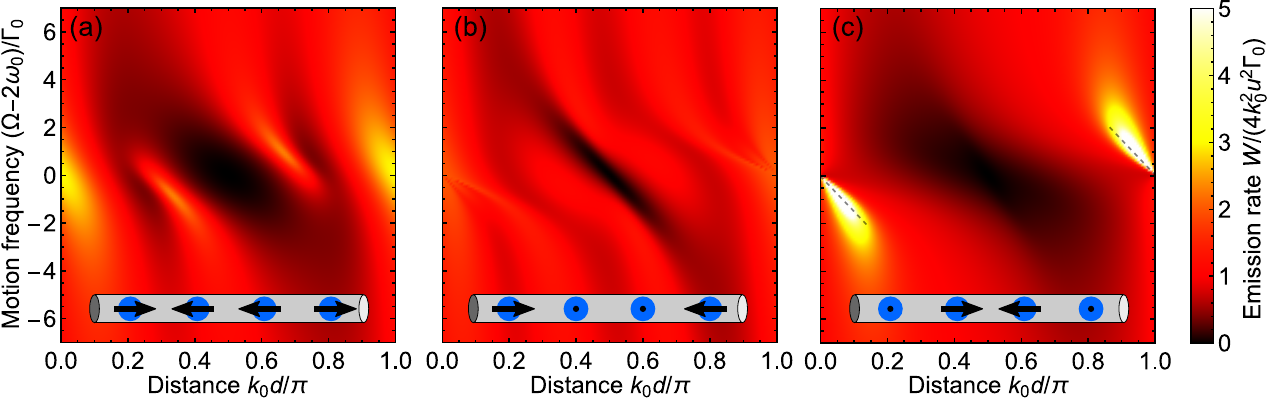}
    \caption{
    The total photon emission rate for the four qubits, when their motion is described by one of the three nontrivial mechanical modes: (a) $u_{o1}$, (b) $u_{e1}$, and (c) $u_{e2}$, as defined in Eqs.~\eqref{eq:modes4}. The schematic representations of the modes are included in the insets. The calculation is based on Eqs.~\eqref{eq:WN} and~\eqref{eq:sigNeig}. Dashed lines in panel (c) indicate the energy of the double-excited dark state, given by Eq.~\eqref{eq:Ed}.
}
    \label{fig:4}
\end{figure*}

A pair of two-level qubits has only one double-excited state, $\psi^{(1)} = (\begin{smallmatrix} 0 & 1 \\ 1 & 0 \end{smallmatrix})/\sqrt{2}$, with energy $E^{(1)} = 2\omega_0 - 2\rmi \Gamma_0$. Consequently, the components of the mechanical self-energy take the form:
\begin{align}\label{eq:S2}
\begin{aligned}
&\Sigma_{12}(\Omega) = \Sigma_{21}(\Omega) = -\frac{k_0^2 \Gamma_0^2 \, \e^{2\rmi k_0 d}}{\Omega - 2\omega_0 + 2\rmi \Gamma_0}, \\
&\Sigma_{11}(\Omega) = \Sigma_{22}(\Omega) = -\rmi k_0^2 \Gamma_0 - \Sigma_{12}(\Omega).
\end{aligned}
\end{align}
The eigenvectors of the matrix $\bm{\Sigma}(\Omega)$ represent the vibrational eigenmodes, each contributing independently to the total emission rate. For the self-energy Eq.~\eqref{eq:S2}, the eigenmodes are $u_{o,e} = (1,\pm1)/\sqrt{2}$, with corresponding self-energy corrections to the mechanical frequencies given by $\Sigma_o = -\rmi k_0^2 \Gamma_0$ and
\begin{align}\label{eq:q2sig}
\Sigma_e(\Omega) = -\rmi k_0^2 \Gamma_0 \frac{\Omega - 2\omega_0 + 2\rmi \Gamma_0 (1 + \e^{2\rmi k_0 d})}{\Omega - 2\omega_0 + 2\rmi \Gamma_0}.
\end{align}

Figure~\ref{fig:2} shows the dependence of the total DCE emission rate induced by the non-trivial $u_e$ vibration, calculated using Eq.~\eqref{eq:q2sig}, as a function of the vibration frequency $\Omega$ and the distance between the qubits. At resonance $\Omega = 2\omega_0$, the emission rate can be enhanced by up to a factor of two compared to a pair of independent qubits. The maximum enhancement occurs for $k_0 d = 0$ or $\pi$. Conversely, for $k_0 d = \pi/2$, the emission vanishes when $\Omega = 2\omega_0$, indicating that under these conditions, the $u_e$ vibration decouples from the light.

\subsubsection{Four qubits}

The emission for an array of three vibrating qubits is similar to that of an array with two qubits. Qualitatively new features arise in the emission of four-qubit arrays, as they can host dark double-excited states~\cite{Ke2019,Poshakinskiy2021dimer}. The dependence of the DCE emission rate induced by three different non-trivial vibrations of the four-qubit array,
\begin{align}\label{eq:modes4}
&u_{o1} = \tfrac{1}{\sqrt4}(1,-1,-1,1) ,\\\nonumber
&u_{e1} = \tfrac{1}{\sqrt2}(1,0,0,-1) , \qquad
u_{e2} = \tfrac{1}{\sqrt2}(0,1,-1,0) ,
\end{align}
is shown in Fig.~\ref{fig:4}. Note that while $u_{o1}$ is the eigenvector of the mechanical self-energy matrix $\bm \Sigma(\Omega)$, the vibrations $u_{e1}$ and $u_{e2}$ are not, in the case of arbitrary $\Omega$. The particular choice of $u_{e1}$ and $u_{e2}$ is determined by the following: Note that the emission from $u_{e2}$, Fig.~\ref{fig:4}(c), is strongly enhanced near $k_0d = 0$ or $\pi$ and $\Omega = 2\omega_0$, while for the other vibrations, it remains finite, see Fig.~\ref{fig:4}(a,b). The reason for the enhancement is the resonance with the double-excited dark state, which has the energy~\cite{Ke2019}
\begin{align}\label{eq:Ed}
E^{\rm (dark)} \approx 2\omega_0 - \frac{14}{3}  k_0 d\,  \Gamma_0 - \frac{157}{27} \rmi k_0^2 d^2 \Gamma_0 \,,
\end{align}
shown by the dashed line in Fig.~\ref{fig:4}(c), and is described by the wave function 
\begin{align}\label{eq:psid}
\psi^{\rm (dark)} = \tfrac1{\sqrt{24}}\left( \begin{smallmatrix}
0 & 2 & -1 & -1 \\
2 & 0 & -1 & -1 \\
-1 & -1 & 0 & 2 \\
-1 & -1 & 2 & 0
 \end{smallmatrix} \right) 
\end{align} 
in the limit $k_0d \to 0$. The coupling of the mechanical vibrations to the dark state is given by the coefficient
\begin{align}
\alpha^{\rm (dark)} = \sqrt{\tfrac{2}{3}}\  (0,1,-1,0) \,,
\end{align}
which explains why it is excited by the $u_{e2}$ vibration only. There also exists another double-excited dark state in the four-qubit array~\cite{Ke2019}. For this state, the calculation yields a zero coupling coefficient, so it is not excited by DCE. 

When the motion frequency matches the energy of the dark state, $\Omega = {\rm Re\,} E^{\rm (dark)}$, see dashed lines in Fig.~\ref{fig:4}(c), the DCE emission is enhanced up to
\begin{align}\label{eq:We2}
W_{e2}^{\rm (max)} = \frac{9}{157}\, \left(\frac{u}{d} \right)^2 \Gamma_0 \,,
\end{align}
where $u$ is the driving amplitude of the mode $u_{e2}$. The emission rate~\eqref{eq:We2} tends to infinity when $k_0 d \to 0$ (and similarly for $k_0 d \to \pi$). However, the large emission leads to a large radiation recoil force on the mechanical motion, which shall lead to a renormalization of the mechanical spectrum, as will be shown in Sec.~\ref{sec:M}, and a finite value of the emission rate. 

The key reason why the dark double-excited modes do get excited by the vibrations, while they cannot be excited by light, is the spatially anti-symmetric wave function of the photon pair emitted by a single moving qubit, Eq.~\eqref{eq:psi1}. It leads to the factor sign$(z_n-z_m)$ in the coupling coefficients $\alpha_i^{(\nu)}$, Eq.~\eqref{eq:alpha}, which makes them non-zero for dark states. Would the qubits be moving perpendicular to the waveguide, or the qubit resonance energy would be modulated as in~\cite{Vyatkin2023}, the excitation of dark states cannot happen. Indeed, in that case, the couplings would be given by $\alpha^{(\nu)}_n = \sum_{m } \psi^{(\nu)}_{nm} \e^{\rmi k_z |z_n - z_m|}$, which vanishes for any double-excited dark state~\cite{Ke2019}.  

Finally, we note that for all three nontrivial vibrations shown in Fig.~\ref{fig:4}, the emission vanishes at $k_0d =\pi/2$, $\Omega=2\omega_0$. This appears to be a general feature for all periodic arrays driven by any kind of vibration that does not involve motion as a whole.


\section{Unruh Effect: Master Equation Approach}\label{sec:Unruh}

The goal of this section is to describe the dynamics and determine the steady state for an array of qubits in a waveguide, in the case of the qubit motion with large amplitudes. Conceptually, the phenomenon considered here is similar to the Unruh effect. However, since the acceleration varies harmonically in time, the effect is resonantly enhanced.

\subsection{Master Equation}\label{sec:mast}

We start with the interaction Hamiltonian Eq.~\eqref{eq:Vrwa}, which we rewrite in the frame rotating with frequency $\Omega/2$ as 
\begin{align}\label{eq:VrwaB}
V_{\rm RWA} = \sum_{k,n} \left[ B_{n,k}^\dag  a_k \e^{\rmi k z_n} + B_{n,k} a_k^\dag \e^{-\rmi k z_n} \right] \,,
\end{align}
where 
\begin{align}
B_{n,k} = g_{n,k}^{(0)} b_n + g_{n,k}^{(1)} b_n^\dag 
\end{align}
is the qubit jump operator corresponding to the emission of a photon with wave vector $k$. Next, we use second-order perturbation theory and the Markovian approximation, following a standard procedure~\cite{Carmichael1993book} to derive the master equation for the density matrix of the qubits,
\begin{align}\label{eq:drhoL}
\frac{d\rho}{dt} =  -\rmi (H_{\rm eff} \rho - \rho H_{\rm eff}^\dag) + J[\rho] + J^\dag[\rho] \equiv \mathcal{L}[\rho]  \,,
\end{align}
where the effective Hamiltonian is given by 
\begin{align}\label{eq:Heff}
H_{\rm eff} &= \sum_n \left(\omega_0 - \tfrac{\Omega}{2}\right) b_n^\dag b_n  \\ \nonumber
&- \frac{\rmi}{c} \sum_n \left( B_{n,+k_0}^\dag B_{n,+k_0} +  B_{n,-k_0}^\dag  B_{n,-k_0} \right) \\ \nonumber
&- \frac{\rmi}{c}  \sum_{n \neq m} \e^{\rmi k_0 |z_n - z_m|}  B_{n,{\rm sign}(z_n - z_m)k_0}^\dag B_{m,{\rm sign}(z_n - z_m)k_0}  \,.
\end{align}
and the stochastic part of the evolution is described by the quantum jump superoperator
\begin{align}\label{eq:J}
 J[\rho] &= \frac1{2c} \sum_{n} 
 \left(B_{n,+k_0} \rho  B_{n,+k_0}^\dag +  B_{n,-k_0} \rho  B_{n,-k_0}^\dag\right) \\\nonumber
 +& \frac1c \sum_{n \neq m} 
 \e^{\rmi k_0 |z_n-z_m|} 
 B_{m,{\rm sign}(z_n-z_m)k_0} \rho  B_{n,{\rm sign}(z_n-z_m)k_0}^\dag  .
 \end{align}
The same result can be obtained using a more universal approach by integrating the photons out from the Keldysh action of the system; see Appendix~\ref{app:Keld}. The latter approach can also be easily generalized to describe more complex types of modulation, e.g., bichromatic modulation.

The system has a weak symmetry~\cite{Albert2014} associated with the parity of the excitation number
\begin{align}
P =  (-1)^{ \sum_n b_n^\dag b_n} \,.
\end{align}
Indeed, $PH_{\rm eff}P = H_{\rm eff}$, and for the jump operators, $PB_{n,k}P = -B_{n,k}$, so $L[P\rho P] = PL[\rho]P$. However, ${\rm Tr} (\rho P)$ is not a conserved quantity. Nevertheless, this implies that the two subspaces of $\rho$ defined by the projections $\rho_{\pm} = (\rho \pm P\rho  P )/2$ are not mixed during the evolution, $d\rho_\pm /dt = L[\rho_\pm]$. If initially all qubits are in their ground states, then $\rho_- = 0$ at all times. Another symmetry of the system originates from the replacements
\begin{align}\label{eq:u1u}
   g_{n,k}^{(0)} \leftrightarrow g_{n,k}^{(1)} \,, \qquad \Delta \to -\Delta\,, \qquad |0\rangle \leftrightarrow |1\rangle \,,
\end{align}
which leave the master equation invariant.

The solution of the master equation describes not only the dynamics and steady state of the system but also allows for determining the emission of the system in the non-equilibrium conditions, which is beyond the perturbative approach of Sec.~\ref{sec:DCEperp}. The explicit expressions for the emission spectrum, power, directivity, and squeezing degree are given in Appendix~\ref{app:Emi}.

In what follows, we will focus on the case of atomic motion perpendicular to the waveguide, as it allows us to obtain certain analytical results even for a large number of atoms. The coupling parameters are taken in the form
\begin{align}
g^{(0)}_{n,k} = g_0 \,, \qquad \qquad  g^{(1)}_{n,k} = v_n g_0 \,,
\end{align}
where we introduced the dimensionless motion amplitude $v_n$. In this case, the master equation simplifies since the jump operators $B_{n,k} = g_0 (b_n + v_n b_n^\dag)$ become $k$-independent, and all ${\rm sign}$ functions in Eqs.~\eqref{eq:Heff} and \eqref{eq:J} can be omitted. The case of atomic motion along the waveguide is considered in Appendix~\ref{app:1mov}.

\subsection{Single qubit}\label{sec:Mod1}

For a single qubit, the density matrix $\rho = 1/2 + \bm{s} \cdot \bm{\sigma}$ can be described by the spin vector $\bm{s}$. The dynamical equations for the qubit spin components in the case of motion perpendicular to the waveguide read:
\begin{align} \label{eq:ds1}
\frac{ds_x}{dt} &= -\Gamma_0 (1-v)^2 s_x + \Delta\, s_y \nonumber \,,\\
\frac{ds_y}{dt} &= -\Gamma_0 (1+v)^2 s_y - \Delta\, s_x \,,\\
\frac{ds_z}{dt} &= -2\Gamma_0(1+v^2) s_z  - \Gamma_0(1-v^2) \,,\nonumber
\end{align}
where $\Delta= \Omega/2-\omega_0$ is the detuning from the parametric resonance, $\Gamma_0 = g_0^2/c$, and we assume that the modulation amplitude $v \equiv g^{(1)}_{1,k}/g_0 $ is real without loss of generality. The steady state of the qubit is described by a diagonal density matrix, with $s_x = s_y = 0$, and the population of the excited state given by
\begin{align}\label{eq:nu1}
n \equiv s_z + \frac{1}{2} =  \frac{v^2}{1+v^2} \,.
\end{align}
The appearance of a finite excited state population of a moving atom can be viewed as a resonant analog of the Unruh effect. In contrast to the original Unruh proposal, we consider periodic motion at a frequency that corresponds to parametric resonance, which greatly amplifies the amplitude of the effect. Note that $n \to 1$ for large modulation amplitudes $v  \gg 1$, in contrast to the case when the system is driven by coherent light, where $n$ is limited to $1/2$~\cite{Scully1997book}.

We also analyze the photon emission characteristics, calculated according to Appendix~\ref{app:Emi}. The photon emission spectrum is the same for both directions $\sigma = \pm$ and is given by
\begin{align}\label{eq:Iu1}
I_{\sigma}(\omega) = \frac{8v^2 \Gamma_0^2 (\Delta_v^2 + \Gamma_v^2)}{[(\omega - \frac{\Omega}{2} + \Delta_v)^2 + \Gamma_v^2][(\omega - \frac{\Omega}{2} - \Delta_v)^2 + \Gamma_v^2]}
\end{align}
where $\omega$ is the emitted photon frequency, $\Delta_v = \sqrt{\Delta^2 - 4v^2\Gamma_0^2}$, and $\Gamma_v = (1+v^2)\Gamma_0$. At small $v$, the spectrum features two peaks at frequencies $\omega_0$ and $\Omega - \omega_0$. As $v$ increases, the peaks broaden and merge at $v \sim \Delta/2\Gamma_0$. The total emission rate
\begin{align}\label{eq:Wu1}
W_{\sigma} \equiv \int I_{\sigma}(\omega) \, \frac{d\omega}{2\pi}= \frac{2v^2\Gamma_0}{1+v^2}
\end{align}
demonstrates no resonant dependence on the modulation frequency $\Omega$, similarly to the consideration in Sec.~\ref{sec:Low1}.

\begin{figure}[t]
\includegraphics[width=0.8\columnwidth]{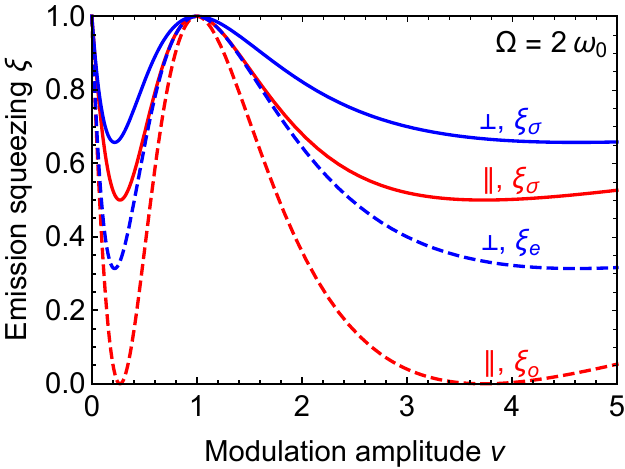}
\caption{Squeezing parameter for electric field at the central frequency emitted  by a single qubit moving with respect to a waveguide. Squeezing of the field emitted in a certain direction corresponds to $\xi_{\sigma}$, while $\xi_{e}$ and $\xi_{o}$ represent the squeezing of the modes that are even and odd with respect to the qubit position, respectively. Calculation is performed for the case of qubit motion along ($\parallel$) and ($\perp$) to the waveguide, using Eqs.~\eqref{eq:Xi1} and~\eqref{eq:Xi1um} for the resonant condition $\Omega=2\omega_0$.}
\label{fig:Xi1}
\end{figure}

The non-classical statistics of the emitted photons are described by the second-order correlation function 
\begin{align}\label{eq:g2u1}
g_{\sigma}^{(2)}(t) = 1 +\frac{(1-v^2)^2}{4v^2} \, \e^{-2\Gamma_v |t|} .
\end{align}
Notice that $g_{\sigma}^{(2)}(t)  > 1$ and tends to infinity as $1/v^2$ for small $v$, indicating that the photons are emitted in pairs. Another characteristic feature of the emitted light is the squeezing of the electric field quadrature. Under resonant conditions, $\Omega = 2\omega_0$, the squeezing of the field emitted in a certain direction at the central frequency $\omega = \Omega/2$ is characterized by the parameter
\begin{align}\label{eq:Xi1}
\xi_{\sigma} = 1 -  \frac{4 v (1-v)^2}{(1+v^2)(1+v)^2} \,.
\end{align}
Figure~\ref{fig:Xi1} shows by the solid blue curve the dependence of $\xi_{\sigma}$ on $v$, calculated from Eq.~\eqref{eq:Xi1}. Note that the dependence is invariant under the transformation $v \to 1/v$, which is a consequence of the  symmetry Eq.~\eqref{eq:u1u}. The strongest squeezing $\xi_{\sigma} = 4\sqrt{2} - 5 \approx 0.66$ is achieved at $v=1+\sqrt{2} \pm \sqrt{2(1+\sqrt{2})} \approx 0.22$ and $4.6$. In the considered here case of motion perpendicular to the waveguide, the emitted field is even with respect to the qubit position. Therefore, it is natural to consider squeezing for the even electric field mode $\xi_{e}$, which is shown by the blue dashed line in Fig.~\ref{fig:Xi1} and is described by the same Eq.~\eqref{eq:Xi1} where the second term should be doubled. Finally, we also plot in Fig.~\ref{fig:Xi1} the squeezing parameters for emission in a certain direction $\xi_\sigma$ (solid red line) and into the odd mode $\xi_{o}$ (dashed red line) for a qubit moving parallel to the waveguide; see Appendix~\ref{app:1mov}. Note that in the latter case, $\xi_{o}$ tends to zero at $v=2\pm \sqrt{3}$, predicting perfect squeezing.

\subsection{Two qubits}

Now, we consider a pair of qubits separated by a distance $d = z_2 - z_1$ and moving perpendicular to the waveguide with amplitudes $v_{1(2)} \equiv g^{(1)}_{1(2),k}/g_0$. We analyze their steady state, which is shown to be entangled, and compute the emission characteristics, which can be directional.

\subsubsection{Unruh entanglement}

  \begin{figure}[tb]
\includegraphics[width=0.9\columnwidth]{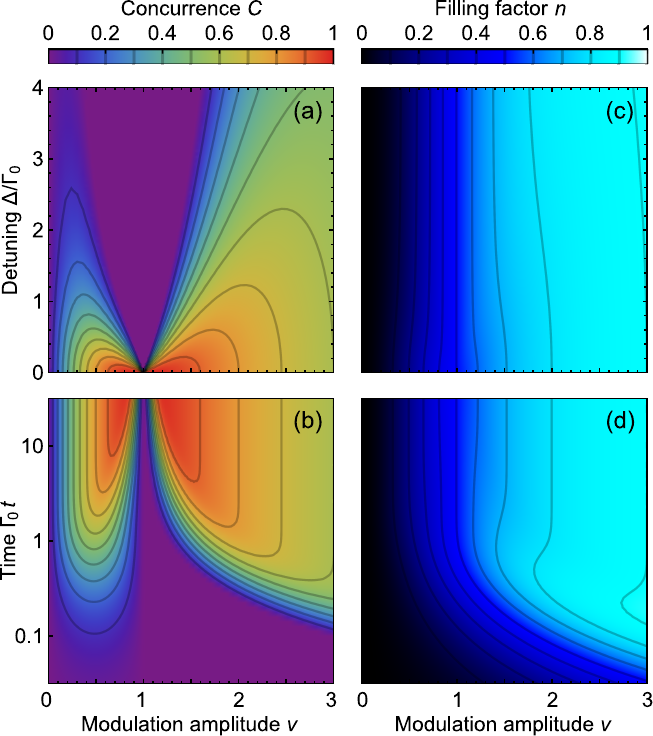}
\caption{
(a,b) The concurrence and (c,d) the average qubit filling factor for a pair of qubits moving perpendicular to the waveguide with equal amplitudes $v_1 = v_2 = v$, calculated for $d = 0$ (mod $\pi/k_0$). Panels (a,c) show the maps of the concurrence and filling factor in the steady state, while panels (b,d) show their temporal evolution for $\Delta = 0$ as the steady state is reached, starting from the ground state $|00\rangle$.
}\label{fig:CNudt}
\end{figure}

 \begin{figure}[tb]
\includegraphics[width=0.9\columnwidth]{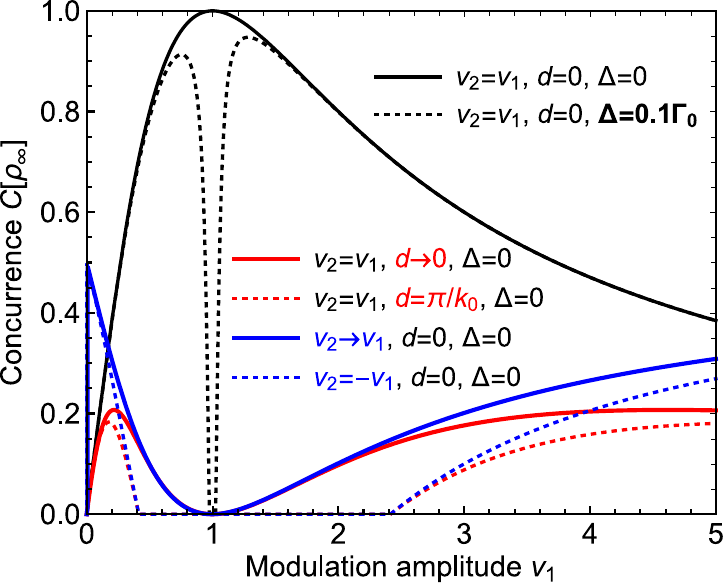}
\caption{
The limiting value of concurrence for a pair of qubits moving along the waveguide, reached at $t \to \infty$ starting from the ground state $|00\rangle$, as a function of modulation amplitude, calculated for various parameters indicated in the graph.
}\label{fig:Cp}
\end{figure}

We begin with the simplest case where the qubits are located at the same coordinate along the waveguide (or separated by an integer number of half-wavelengths), i.e., $d = 0$ (mod $\pi/k_0$), and exhibit identical motion, $v_1 = v_2 = v$. In this case, the master equation simplifies to
\begin{align}\label{eq:drhoS}
&\frac{d\rho}{dt} = \rmi \Delta [ S_z , \rho ] - \Gamma_0 (\rho S_v^\dag S_v +  S_v^\dag S_v \rho - 2 S_v \rho S_v^\dag ),
\end{align}
with a collective jump operator
\begin{align}\label{eq:rhoSv}
S_v = (S_x - \rmi S_y) + v  (S_x + \rmi S_y),
\end{align}
where we introduced the total spin operator $\bm S = \bm S_1 + \bm S_2$, with $S_{n,x} = (b_n + b_n^\dag)/2$, $S_{n,y} = \rmi (b_n - b_n^\dag)/2$, and $S_{n,z} = (b_n^\dag b_n - 1 )/2$. Such a system possesses a strong symmetry~\cite{Albert2014} associated with the absolute value of the total spin. Indeed, the operator $\bm S^2$ commutes with both the Hamiltonian and the jump operator. Therefore, ${\rm Tr} (\bm S^2 \rho)$ is conserved. The other symmetry Eq.~\eqref{eq:u1u} yields that the dynamical equation~\eqref{eq:drhoS} is invariant under the transformation 
\begin{align}\label{eq:v1v2}
&v \to 1/v\,, \quad \Delta \to \Delta/v^2\,, \\ \nonumber
&\rho(t) \to \e^{\rmi\pi S_x}\rho^*(v^2 t)\, \e^{-\rmi\pi S_x} \,,  
\end{align}
where the operator $\e^{\rmi\pi S_x}$ does the interchange $|0\rangle \leftrightarrow |1\rangle$.

The master equation~\eqref{eq:drhoS} has two steady states. First, there is the pure dark state
\begin{align}\label{eq:psiS0}
|\psi_{d}\rangle = \frac{|01\rangle - |10\rangle}{\sqrt{2}},
\end{align}
which has a total spin $S=0$, so $S_v |\psi_d\rangle = S_v^\dag |\psi_d\rangle = 0$. However, this state cannot be reached starting from any state other than itself because all other states have a total spin $S=1$. If we start from one of the states with $S=1$, in particular from the ground state $|00\rangle$, the other steady state is reached:
 \begin{align}\label{eq:rhoS1}
&\rho_{\infty} = \tfrac1{(1-v^4)^2 + \Delta^2 (1+v^2+v^4)} \\\nonumber
& \times\left(  \begin{smallmatrix} v^2(1-v^2)^2(1+v^2)+\Delta^2 v^4 & 0& 0& -v(1-v^2)^2(1+v^2 +\rmi \Delta)\\ 0& \frac{v^2\Delta^2}2 & \frac{v^2\Delta^2}2 & 0\\ 0& \frac{v^2\Delta^2}2 & \frac{v^2\Delta^2}2 & 0\\ -v(1-v^2)^2(1+v^2 -\rmi \Delta) &0 & 0 & (1-v^2)^2(1+v^2)+\Delta^2 \end{smallmatrix} \right) \,,
 \end{align}
where we used the basis $|00\rangle, |01\rangle, |10\rangle, |11\rangle$.

Notably, the steady state described by Eq.~\eqref{eq:psiS0} is, in fact, the entangled Bell state, and the other steady state given by Eq.~\eqref{eq:rhoS1} can also possesses entanglement. In Fig.~\ref{fig:CNudt}(a), we show the calculated dependence of the two-qubit concurrence $C$ for the steady state $\rho_{\infty}$, Eq.~\eqref{eq:rhoS1}, on the modulation amplitude $v$ and the detuning $\Delta$. The concurrence is strongest at zero detuning, $\Delta = 0$, and decreases with increasing $|\Delta|$, especially in the region $v \sim 1$. In fact, at $\Delta=0$, the state in Eq.~\eqref{eq:rhoS1} becomes a pure entangled state with the wave function
\begin{align}
|\psi_{\infty}\rangle = \frac{|00\rangle - v |11\rangle}{\sqrt{1+v^2}},
\end{align}
and concurrence $C = 2v/(1+v^2)$. The maximal concurrence $C=1$ is achieved at $v \to 1$, when $|\psi_\infty\rangle$ becomes a Bell state. However, note that for $v \to 1$, the concurrence is particularly sensitive to even small detuning.

Figure~\ref{fig:CNudt}(b) shows the time evolution of the concurrence, starting from the separable ground state $|00\rangle$, calculated for $\Delta=0$. Interestingly, for $v=1$, the concurrence remains zero at all times. However, in the vicinity of $v=1$, the concurrence can approach values arbitrarily close to 1, provided that one waits long enough. For comparison, we also plot in Fig.~\ref{fig:CNudt}(c,d) the dependence of the qubit filling factor $n= {\rm Tr}(b_1^\dag b_1 \rho)$ on the modulation amplitude, detuning, and time. Unlike the concurrence, the filling factor shows almost no dependence on $\Delta$ and does not exhibit any peculiarities as $v \to 1$, cf. Figs.~\ref{fig:CNudt}(a),(b) and~(c),(d).

Even a small deviation from the conditions $v_1 = v_2$ and $d = 0$ can lead to an abrupt change in the limiting steady state $\rho_\infty$ that is reached at infinite time starting from the ground state $|00\rangle$. This is due to the breaking of the symmetry associated with $\bm S^2$ conservation, reducing the steady-state subspace from two states to just one. Figure~\ref{fig:Cp} shows the concurrence of the reached steady state $C[\rho_\infty]$ for various parameters. The solid black curve correspond to the already discussed conditions $v_1 = v_2$, $d = 0$, and $\Delta=0$. The other curves show the results when any of these three conditions are violated. In all cases, the limiting value of concurrence is quenched in the region around $v \sim 1$. If $v_1 \neq v_2$ or $d \neq 0$, the limiting value of concurrence is bounded by $C[\rho_\infty] \leq 0.5$. However, at finite times, the concurrence can still reach larger values.

Interestingly, if $d = 0$ but $v_1 \neq v_2$ (blue curves in Fig.~\ref{fig:Cp}), the limiting steady-state density matrix $\rho_\infty$ is non-trivial even in the limit of vanishing modulation amplitude $v_1, v_2 \to 0$. This is due to the possibility of exciting the dark state $|\psi_d\rangle$, Eq.~\eqref{eq:psiS0}, which has an infinite lifetime at $d = 0$. The steady state of the system reads
\begin{align}
\rho_{\infty} = \frac{|00\rangle\langle 00| + |\psi_d\rangle \langle\psi_d| }{2},
\end{align}
and is characterized by a concurrence $C=0.5$.

To summarize, the emergent entanglement of the moving qubits, which can be termed Unruh entanglement, is highly sensitive to the system parameters. Strong entanglement can be achieved with proper tuning of the parameters and sufficient waiting time.

 \subsubsection{Entangled directional DCE emission}
 
 \begin{figure*}[tb]
\includegraphics[width=0.9\textwidth]{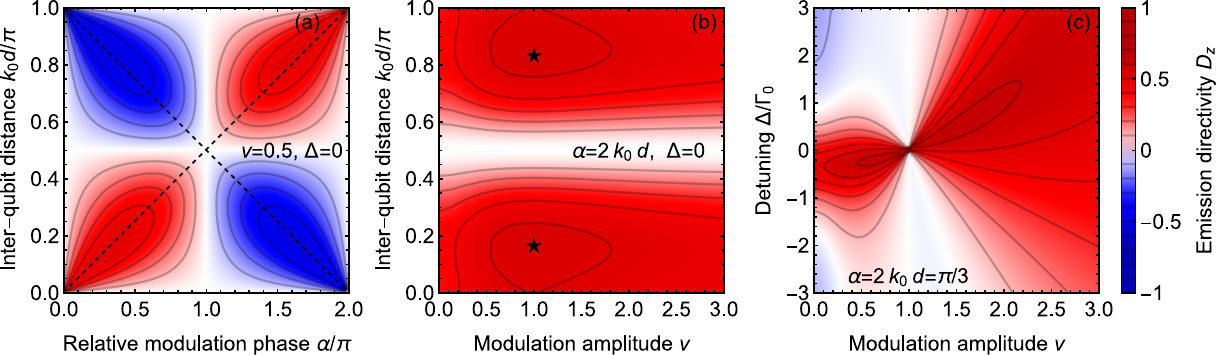}
\caption{
The directivity $D_z$ of photons emitted by a pair of qubits moving perpendicular to the waveguide with a phase difference $\alpha$. The parameters are indicated in the graph. Dashed lines in panel (a) represent the constructive interference condition for photon pairs emitted in a certain direction. Stars in panel (b) denote the maximum value of $D_z$ as a function of $\alpha$, $d$, and $v$ at zero detuning $\Delta=0$. 
}\label{fig:Dir}
\end{figure*}
 \begin{figure*}[tb]
\includegraphics[width=0.9\textwidth]{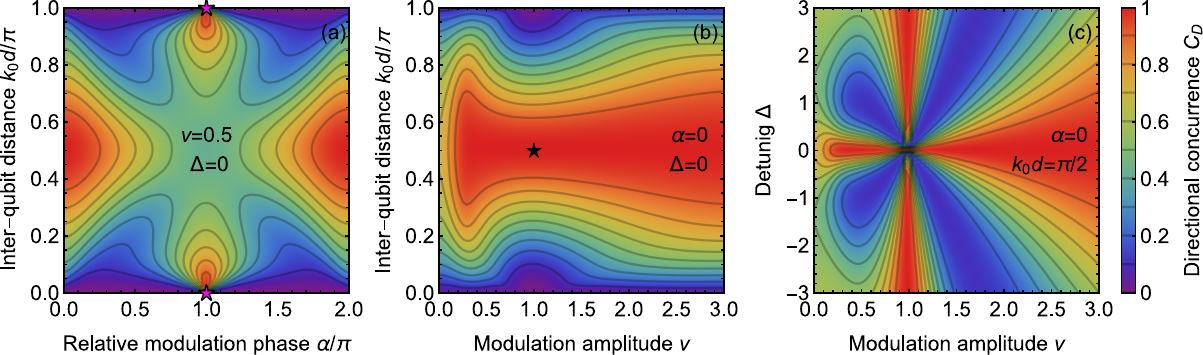}
\caption{
The two-photon directional concurrence $C_D$ for the light emitted by a pair of qubits moving perpendicular to the waveguide with a phase difference $\alpha$. The parameters are indicated in the graph. The stars in panels~(a) and~(b) denote the parameters at which maximally entangled Bell states with $C_D=1$ are emitted.
}\label{fig:CDir}
\end{figure*}

Now, we describe the emission properties of a pair of qubits moving perpendicularly to the waveguide. Similar to the single-qubit case, the emitted light is strongly non-classical, as evidenced by the second-order correlations and field quadrature squeezing. However, our focus here is on the emerging directivity of the emission, which was absent for a single qubit. To achieve directivity, the spatial parity must be broken, so the qubits must be located at different points, $d \neq 0$, and move differently, $v_1 \neq v_2$. We choose $v_{1,2} = v\, \e^{\mp \rmi \alpha/2}$, where $\alpha \neq 0$ is the phase delay between the motions of the qubits.

To characterize the directivity of the emitted photons, we use the directional density matrix
\begin{align}\label{eq:rho1p}
\rho^{(1)} =  \frac{J^{(1)}}{\text{Tr\,} J^{(1)}}, \quad
J_{\sigma\sigma'}^{(1)}(t) = \langle a_{\sigma'}^{{\rm out}\,\dag}(t) a_{\sigma}^{\rm out}(t) \rangle
\end{align}
where $a_{\sigma}^{\rm out}(t)$ is the annihilation field operator for the emitted photons traveling in the direction $\sigma$; see Appendix~\ref{app:Emi} for details. The directivity of emission is calculated as 
\begin{align}
D_z = \text{Tr\,} \sigma_z \rho^{(1)} \,,
\end{align}
where $D_z = \pm 1$ indicates that all photons are emitted in the direction $\pm z$. Other components of the directional density matrix correspond to photons being emitted in modes with even or odd spatial parity.

Figure~\ref{fig:Dir} shows the calculated directivity $D_z$ as a color map. Panel (a) displays the dependence of $D_z$ on the relative modulation phase $\alpha$ and the inter-qubit distance $d$. As expected, the directivity vanishes at $k_0 d = 0, \pi/2, \pi$ or $\alpha = 0,\pi,2\pi$, where the system restores certain parity. The directivity is maximal when $\alpha = \pm 2 k_0 d$, as indicated by the dashed lines in Fig.~\ref{fig:Dir}(a). In this condition, the pairs of photons emitted by the first and second qubits in the direction $\pm z$ interfere constructively. In Fig.~\ref{fig:Dir}(b), we fix the relation $\alpha = 2 k_0 d$, where forward scattering is enhanced, and examine the dependence of $D_z$ on the modulation amplitude $v$. The directivity is significant even at small $v$, but the maximal value $D_z = 9/17 \approx 0.53$ is achieved at $\alpha = 2 k_0 d = \pi/3$ and $u \to 1$, as indicated by a star in Fig.~\ref{fig:Dir}(b). The plots described above were calculated for zero detuning $\Delta = 0$. The dependence of $D_z$ on detuning and modulation amplitude is shown in Fig.~\ref{fig:Dir}(c). It is evident that at $v \to 1$, the directivity can be significantly affected by even a small detuning. For $v \neq 1$, introducing a detuning can lead to an increase in directivity beyond the maximum value of $0.53$ described above, though this comes with a reduction in emission intensity.

The directional density matrix defined above, Eq.~\eqref{eq:rho1p}, characterizes the propagation direction of a single emitted photon. Since the photons in the DCE are emitted in pairs, it is natural to expect that their emission directions are correlated. To describe this correlation, we first calculate the two-photon wave function 
\begin{align}\label{eq:psi2u}
\psi_{\sigma_1\sigma_2}(t_1,t_2) = \langle a_{\sigma_1}^{\rm out}(t_1) a_{\sigma_2}^{\rm out}(t_2)\rangle \,,
\end{align}
which has a trivial dependence on $t_1+t_2$, i.e., $\psi_{\sigma_1\sigma_2}(t_1,t_2) = \psi_{\sigma_1\sigma_2}(t_1-t_2) \, \e^{-\rmi\Omega(t_1+t_2)/2}$. Next, we introduce the time-averaged two-photon directional density matrix, defined as $\rho^{(2)} =  J^{(2)}/\text{Tr\,} J^{(2)}$, where
\begin{align}\label{eq:J2u}
J_{\sigma_1\sigma_2,\sigma'_1\sigma_2'}^{(2)} = \int_0^\infty \psi_{\sigma_1\sigma_2}(\tau)  \psi_{\sigma_1'\sigma_2'}^*(\tau) \, d\tau \,.
\end{align}
An alternative way to define a two-photon density matrix is via $\langle a_{\sigma_2'}^{\rm out}(0) a_{\sigma_1'}^{\rm out}(\tau) a_{\sigma_1}^{\rm out}(\tau) a_{\sigma_2}^{\rm out}(0)\rangle$. However, this density matrix includes not only the coherent photons generated in the same parametric generation event but also photons emitted in different events. Averaging over $\tau$ would then be influenced by the contribution of the latter incoherent photons, which is finite even at $\tau \to \infty$, resulting in a separable form of the corresponding density matrix. Thus, we adhere to the definition given in Eqs.~\eqref{eq:psi2u} and \eqref{eq:J2u}, where the coherent part is explicitly filtered.

To characterize the quantum correlations between the directions of photon emission, we calculate the directional concurrence $C_{D}$ corresponding to $\rho^{(2)}$. Figure~\ref{fig:CDir} shows the color maps of $C_D$. Comparing the dependence of directional concurrence on the relative modulation phase $\alpha$ and the inter-qubit distance $d$ in Fig.~\ref{fig:CDir}(a) with the corresponding dependence of directivity in Fig.~\ref{fig:Dir}(a), one can see that concurrence and directivity are mutually exclusive. Indeed, if photons are emitted all in one direction, the propagation directions of a pair of photons cannot be entangled. Conversely, if photon pairs are directionally entangled, the individual photons must be emitted in various directions.

The directional entanglement is maximal in the following two cases: 
(i) At $k_0d = \pi/2$ and $\alpha = 0$, the processes in which photon pairs are emitted by the two qubits in the same direction, i.e., $|++\rangle$ or $|--\rangle$, interfere destructively. The pure Bell state $(|+-\rangle + |-+\rangle)/\sqrt{2}$ with $C_D=1$ is emitted in the limit $v \to 1$, as indicated by the black star in Fig.~\ref{fig:CDir}(b). In the rather broad region around this point, for modulation amplitudes $0.2 \lesssim v \lesssim 5$, the concurrence remains very high, $C_D > 0.9$. 
(ii) At $k_0d \to 0, \pi$ and $\alpha = \pi$, the two-photon state must have odd spatial parity. Consequently, the pure Bell state $(|++\rangle - |--\rangle)/\sqrt{2}$ is emitted, yielding $C_D=1$ for any value of $v > 0$. This case is marked by magenta stars in Fig.~\ref{fig:CDir}(a).

The above results are valid for zero detuning $\Delta = 0$. Figure~\ref{fig:CDir}(c) shows the dependence of the directional concurrence on $\Delta$. Clearly, detuning can destroy the directional concurrence, which is particularly vulnerable around $v = 1$.

\subsection{Large numbers of qubits}

\begin{figure}[tb]
\includegraphics[width=0.9\columnwidth]{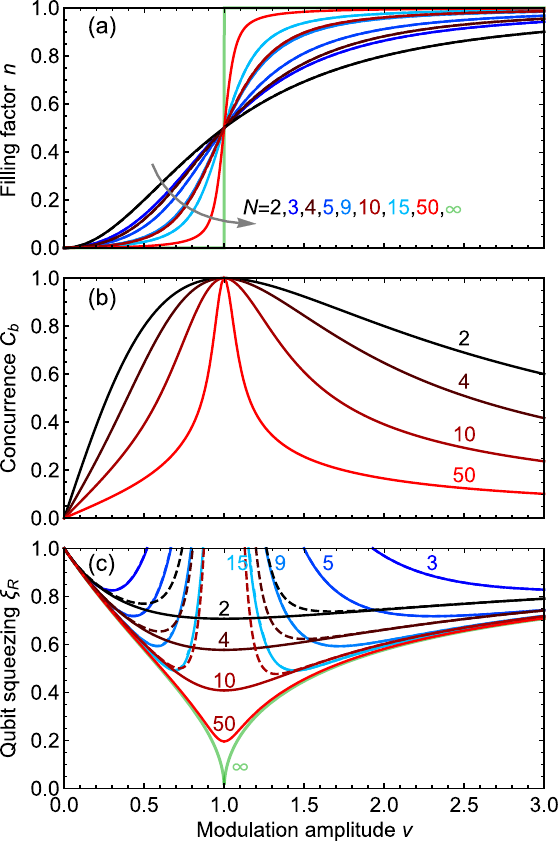}
\caption{(a) Qubit filling factor, (b) concurrence $C_b$ between one qubit and all other qubits, and (c) the squeezing factor for an ensemble of $N$ qubits located at the same point and moving perpendicular to the waveguide with the same amplitude $v$. Solid curves show calculations for zero detuning $\Delta=0$, while dashed curves correspond to $\Delta = 0.1\Gamma_0$. The number of qubits $N$ is indicated in the graph.}\label{fig:PhTr}
\end{figure}

Now we consider a large number of qubits $N$ moving perpendicularly to the waveguide. For simplicity, we limit the discussion to the case where all qubits are located at the same coordinate along the waveguide (or separated by an integer number of half-wavelengths) and move with equal amplitudes: $z_n = 0$ (mod $\pi/k_0$), $v_n  \equiv g^{(1)}_{n,k}/g_0 = v$ for all $n=1, \ldots, N$, where $v$ is assumed to be real. Then, similarly to the case of two qubits, the total spin $\bm{S}^2$ is conserved, and the dynamics are described by Eqs.~\eqref{eq:drhoS} and \eqref{eq:rhoSv}. The transformation Eq.~\eqref{eq:v1v2} indicates that for $\Delta =0 $ and $v =1$, the system has a strong symmetry associated with the conservation of ${\rm Tr}(\rho S_x)$. As shown in Ref.~\cite{Munos2019} for a similar Lindbladian, dissipative phase transitions are expected to occur at this point in the limit $N \to \infty$.

We assume that we start with the ground state $|0\rangle^{N}$, so we must remain in the subspace with $S=N/2$. Interestingly, for even $N$ and $\Delta = 0$, the final steady state is pure and can be found analytically from the condition $S_v|\psi_\infty \rangle = 0$, which yields~\cite{DallaTorre2013}
\begin{align}
|\psi_\infty \rangle \propto \sum_{n=0}^{N/2} (-v)^n \frac{C_{N/2}^n}{\sqrt{C_N^{2n}}} | 2n \rangle
\end{align}
where $| 2n \rangle$ is the normalized symmetric state with $2n$ qubit excitations ($S_z = -N/2 + 2n$). The average filling factor $n = \langle \psi_\infty | b_m^\dag b_m |\psi_\infty \rangle$ then reads
\begin{align}
n = \frac{v^2\ {}_2F_1(\tfrac{3}{2},1-\tfrac{N}{2};\tfrac{3-N}{2};v^2)}{(N-1)\ {}_2F_1(\tfrac{1}{2},-\tfrac{N}{2};\tfrac{1-N}{2};v^2)} \,,
\end{align}
where $_2F_1$ is the hypergeometric function that reduces to a polynomial for the given parameters. For odd $N$, the steady state is mixed, $\rho_\infty \propto (S_v^\dag S_v)^{-1}$. Figure~\ref{fig:PhTr}(a) shows the dependence of the filling factor $n$ on the modulation amplitude for different numbers of qubits $N$. Clearly, for $N \to \infty$, regardless of parity, the dependence converges to the step function $n \to \theta (|v|-1)$, indicating a first-order phase transition. The introduction of a detuning $\Delta$ has virtually no effect on this dependence. From the invariance of the system with respect to the transformation Eq.~\eqref{eq:v1v2}, it follows that $n=1/2$ at $v=1$ for any $\Delta$.

To describe this phase transition, we use the Holstein–Primakoff approximation. We assume $v < 1$, so the system is close to the $S_z = -N/2$ state, and set $S_x - \rmi S_y = \sqrt{N} c$, $S_z = -N/2 + c^\dag c$, where $c$ is a new bosonic annihilation operator. Next, we perform a Bogolyubov transform by introducing another bosonic annihilation operator $d = (c + v c^\dag)/\sqrt{1-v^2}$ and obtain the dynamic equation
\begin{align}\label{eq:drhoPT}
\frac{d\rho}{dt} &= \frac{\Delta}{1-v^2} (d^\dag - v d)(d - v d^\dag) \\ \nonumber
&- (1-v^2) N \Gamma_0 (\rho d^\dag d + d^\dag d \rho - 2 d \rho d^\dag) \,.
\end{align}
We observe that in the limit $N \to \infty$, the detuning $\Delta$ becomes irrelevant and can be ignored. Thus, the system possesses a pure steady state, determined from $d | \psi_\infty \rangle = 0$, which gives the average filling factor
\begin{align}
n \equiv \frac{1}{N} \langle \psi_\infty | c^\dag c | \psi_\infty \rangle = \frac{1}{N} \frac{v^2}{1-v^2} \,.
\end{align}
The Holstein-Primakoff approximation is valid if the filling factor is small, i.e., for $|v-1| \gtrsim 1/N$. Therefore, in the limit $N \to \infty$, we indeed find $n \to 0$ for all $v < 1$. It also follows from the dynamic equation~\eqref{eq:drhoPT} that the asymptotic decay rate towards the steady state is $(1-v^2) N \Gamma_0$. This rate is positive when $v < 1$ and vanishes as $v \to 1$, indicating a first-order dissipative phase transition~\cite{Kessler2012}. For $v > 1$, the true steady state is one with an average filling close to 1, which can be described similarly using the Holstein-Primakoff transform near that point.

To quantify the emergent entanglement of the qubits, we start with the case where their state is pure, i.e., $N$ is even and $\Delta = 0$. We consider the bipartition where one qubit is separated from all the others and calculate the corresponding concurrence $C_b = \sqrt{2(1 - \text{Tr}\,\rho_1^2)}$ with $\rho_1 = \text{Tr}_{2 \ldots N} |\psi_\infty \rangle \langle \psi_\infty |$, see Fig.~\ref{fig:PhTr}(b). Although the concurrence $C_b$ decreases with increasing $N$, it remains strong at $v \to 1$. In this limit, the steady state satisfies $S_x |\psi_\infty \rangle = 0$ and yields $C_b = 1$.

To quantify the entanglement in the general case, where the steady state is mixed, we use the spin squeezing parameter $\xi_R = \sqrt{N \langle S_x^2 \rangle} / |\langle S_z \rangle|$~\cite{Ma2011}. Low values of $\xi_R < 1$ indicate the presence of entanglement~\cite{Sorensen2001nat} and also quantify its depth~\cite{Sorensen2001prl}. Figure~\ref{fig:PhTr}(b) shows, with solid curves, the dependence of $\xi_R$ on the modulation amplitude $v$ for $\Delta = 0$ and different numbers of qubits $N$. Clearly, the parity of $N$ strongly affects the value of the squeezing parameter $\xi_R$. For even $N$ (red curves), the strongest squeezing is achieved at $v \to 1$, while for odd $N$ (blue curves), the squeezing appears to be suppressed around $v = 1$. A similar suppression of squeezing also occurs for even $N$ if a small detuning $\Delta = 0.1\Gamma_0$ is introduced, as shown by the dashed curves. For odd $N$, the detuning has almost no effect on the squeezing, which is already suppressed around $v = 1$.

In any case, as $N$ increases, the squeezing parameter tends to the same limit, shown by the green line in Fig.~\ref{fig:PhTr}(b). However, the rate of convergence is significantly lower for even $N$ or nonzero $\Delta$. The limiting value can be most easily obtained by considering the case of even $N$ and $\Delta = 0$, where from $S_v |\psi_\infty \rangle = 0$ and the canonical commutation relations for the spin operators, we find the exact relation
\begin{align}
\langle \psi_\infty | S_{x,y}^2 |\psi_\infty \rangle = \frac{v \mp 1}{v \pm 1} \frac{\langle \psi_\infty | S_z |\psi_\infty \rangle}{2}\,.
\end{align}
Then, recalling that in the limit $N \to \infty$, $\langle S_z \rangle \to (N/2) \text{sign}(v^2-1)$, we obtain the limiting value of the squeezing parameter
\begin{align}\label{eq:XiR}
\xi_R \to \sqrt{\left|\frac{1-v}{1+v}\right|} \,,
\end{align}
which is reached in the region $|v-1| \gtrsim 1/N$.


\section{Optomechanical backaction}\label{sec:Back}

In this section, we examine the backaction of the DCE emission on the mechanical motion, focusing specifically on the motion along the waveguide. Unlike previous sections, we consider the system without an external mechanical drive. Instead, we assume that the mechanical motion is quantized and described by the free Hamiltonian $H_m = \Omega_0 \sum_n c_n^\dag c_n$, where $c_n$ are the bosonic annihilation operators corresponding to the motion of the $n$-th qubit and $\Omega_0$ is the mechanical eigenfrequency. The optomechanical interaction is given by Eqs.~\eqref{eq:V1} and \eqref{eq:g1mo}, with the motion amplitude in the form
\begin{align}
u_n = u_0 (c_n + c_n^\dag),
\end{align}
where $u_0 = 1 / \sqrt{2M\Omega_0}$ is the zero-point motion amplitude and $M$ is the effective mass. We assume that $k_0 u_0 \lesssim 1$, so the perturbative approach of Sec.~\ref{sec:DCEperp} can be applied. However, we will not limit the calculations to the first order in $k_0 u_0$. Higher-order corrections will be considered in a self-consistent manner to properly describe the optomechanical effects.

\subsection{Modification of the mechanical spectrum by DCE recoil}\label{sec:M}

\begin{figure}
    \centering
    \includegraphics[width=0.75\columnwidth]{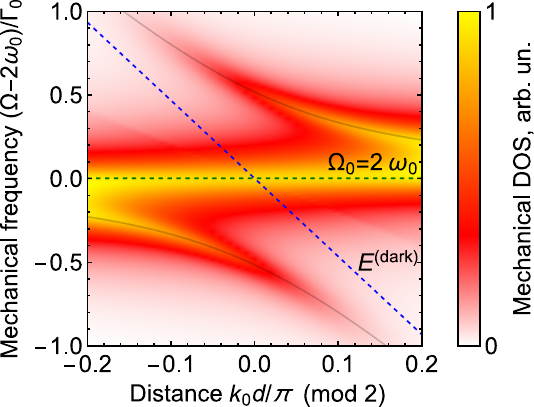}
    \caption{Mechanical density of states ${\rm DOS}(\Omega)$ in the system of four qubits, shown as a function of $\Omega$ and the distance between the qubits. Calculation is performed after Eqs.~\eqref{eq:Gm} and \eqref{eq:DOS} for  $\Omega_0 = 2\omega_0$ and $k_0^2 u_0^2 = 0.1$. Dashed lines indicate the position of the bare mechanical resonance at frequency $\Omega_0$ and the energy of the dark double-excited state, Eq.~\eqref{eq:Ed}. Solid lines represent the eigenenergies of the hybridized polaritonic excitations calculated from the effective Hamiltonian, Eq.~\eqref{eq:He2d}. }
    \label{fig:dos}
\end{figure}

The DCE radiation extracts energy from the mechanical motion, leading to a dissipative radiation friction force. For a single qubit, the radiative decay rate of the amplitude of free mechanical oscillations can be calculated as $\gamma = \hbar \omega_0 W / (2 E_m)$, where $E_m = 2M \Omega_0^2 |u_1|^2$ is the mechanical energy and $W$ is the radiation rate given by Eq.~\eqref{eq:W1}. This yields
\begin{align}
\gamma = \frac{\hbar \omega_0 \Gamma_0}{M c^2},
\end{align}
where we used $\Omega_0 \approx 2 \omega_0$ and restored $\hbar$ for clarity. The same result can be obtained from the self-energy of the mechanical motion in the system. For a single qubit, only the first term is present in Eq.~\eqref{eq:sigNeig}, giving $\Sigma(\Omega) u_0^2 = -\rmi \gamma$.

In the case of a qubit array, the recoil force can be accounted for by considering the Green's function of the qubit vibrations
\begin{align}\label{eq:Gm}
\bm{\mathcal{G}}(\Omega) = \left[ \Omega - \Omega_0 - u_0^2 \bm \Sigma(\Omega) \right]^{-1},
\end{align}
where the self-energy matrix $\bm \Sigma(\Omega)$ is given by Eq.~\eqref{eq:sigNeig}. The eigen vibrations of the system are determined by the eigenvectors of $\bm \Sigma(\Omega)$, and its eigenvalues determine the corrections to the mechanical eigenfrequencies. The corrections have both real and imaginary parts, originating from the conservative and dissipative components of the radiation recoil force. We have already used the dissipative part of the recoil to calculate the total emission rate, see Eq.~\eqref{eq:WN}. The conservative part of the recoil force leads to modifications in the mechanical spectrum, which we will now describe.

To quantify the modification of the mechanical spectrum in the presence of the radiation recoil force, we calculate the mechanical density of states,
\begin{align}\label{eq:DOS}
{\rm DOS}(\Omega) = -\frac{1}{\pi} \, {\rm Im} \, {\rm Tr} \, \bm{\mathcal{G}}(\Omega).
\end{align}
Figure~\ref{fig:dos} shows the dependence of the DOS on the inter-qubit distance for a structure with four qubits. To maximize the effect of the recoil force, the bare mechanical frequency $\Omega_0$ is assumed to fulfil the parametric resonance condition $\Omega_0=2\omega_0$. The central peak of the DOS at $\Omega \approx \Omega_0$ comprises the three mechanical vibrations: the motion as a whole, $u_{o1}$, and $u_{e1}$, see Eq.~\eqref{eq:modes4} for mode definitions. These modes acquire a finite width due to the dissipative part of the radiation recoil force and a small correction to their frequencies due to its conservative part. However, the latter is of the order of the peak width and cannot be resolved.

On the other hand, the $u_{e2}$ mode interacts strongly with the dark double-excited state, Eqs.~\eqref{eq:Ed}-\eqref{eq:psid}, leading to the avoided-crossing pattern in Fig.~\ref{fig:dos} and the formation of hybrid excitations, which we term biphonoritons. As a mixture of two photons and a phonon, a biphonoriton is the quantum counterpart of a phonoriton in semiconductor structures~\cite{Keldysh1982,Hanke1999,Poshakinskiy2017,Poshakinskiy2021,Latini2021,Kuznetsov_2023,Santos2023}. Conventional phonoritons result from the hybridization of a phonon and a polariton (the latter being a mixed half-light half-exciton quasiparticle), and they are observed in spectra under strong coherent optical pumping. In contrast, the biphonoritons considered here can be excited by a pair of photons only, see Sec.~\ref{sec:g2}. The formation of biphonoritonic modes can be described by the effective Hamiltonian
\begin{align}\label{eq:He2d}
H_{\rm e2-dark} = \left( \begin{array}{cc}
\Omega_0 - \rmi k^2 u_0^2 \Gamma_0 & \frac{2 \sqrt{2}}{3} k u_0 \Gamma_0 \\
\frac{2 \sqrt{2}}{3} k u_0 \Gamma_0 & E^{\rm{(dark)}}
\end{array} \right),
\end{align}
where the diagonal terms represent the bare modes, shown by dashed lines in Fig.~\ref{fig:dos}, and the off-diagonal terms describe the strength of their coupling, similar to the vacuum Casimir-Rabi splitting in optomechanical cavities~\cite{Macri2018}. The real parts of the eigenvalues of the Hamiltonian Eq.~\eqref{eq:He2d}, which correspond to the energies of the biphonoritons, are plotted by solid lines in Fig.~\ref{fig:dos} and coincide with the peaks of the DOS. For the avoided crossing shown in Fig.~\ref{fig:dos} to be resolved, i.e., for strong coupling between the mechanical motion and the dark double-excited state, the coupling strength should exceed the decay rates of both modes, leading to the condition $|d \mod \pi/k_0| \ll u_0$.

\subsection{Effect of DCE recoil on correlations of the scattered photons}\label{sec:g2}

We now consider the system excited by low-intensity coherent light and study the correlations of the scattered photons, taking into account the possibility that these photons can excite vibrations of the qubits. A single photon cannot excite a vibration; therefore, single-photon transmission and reflection are described by the same coefficient $t_{\sigma'\sigma}(\omega)$ as in the absence of optomechanical interaction. When two photons are incident on the system simultaneously, they can be converted into a vibration and subsequently emitted as photons with different frequencies. Another mechanism for two-photon interaction is provided by qubit nonlinearity. Both frequency-mixing processes are described by the scattering matrix $M_{\sigma_1'\sigma_2',\sigma_1\sigma_2}(\omega_1',\omega_2'; \omega_1, \omega_2)$~\cite{Yudson1984,Shen2007,Ke2019,Sheremet2023}, where quantities without and with a prime correspond to incident and scattered photons, respectively (see Appendix~\ref{app:Dia} for calculation details). The statistics of the back-scattered light can be described by the second-order correlation function $g^{(2)}(\tau)$, which is given by~\cite{Poshakinskiy2016}
\begin{align}\label{eq:g2M}
&g^{(2)}(\tau)\\\nonumber
&=\left| 1 + \frac{\rmi  \int M_{--,++}(\omega+\omega',\omega-\omega';\omega,\omega) \e^{-\rmi \omega' \tau} d\omega'}{4\pi t_{-+}^2} \right|^2 .
\end{align}
We now apply the above equation to cases of a single qubit and qubit arrays with two or four qubits.

 \begin{figure}
    \centering
    \includegraphics[width=0.75\columnwidth]{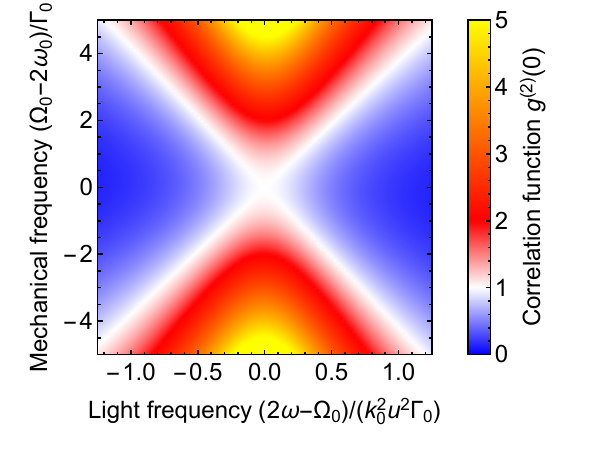}
    \caption{Second-order correlation function of the light reflected from a single qubit that can move along the waveguide, shown as a function of the detunings between the incident light frequency $\omega$, mechanical motion eigenfrequency $\Omega_0$, and qubit resonance frequency $\omega_0$. The calculation was performed using Eq.~\eqref{eq:g2N1}.}
    \label{fig:g1}
\end{figure}

 \begin{figure}
    \centering
    \includegraphics[width=0.99\columnwidth]{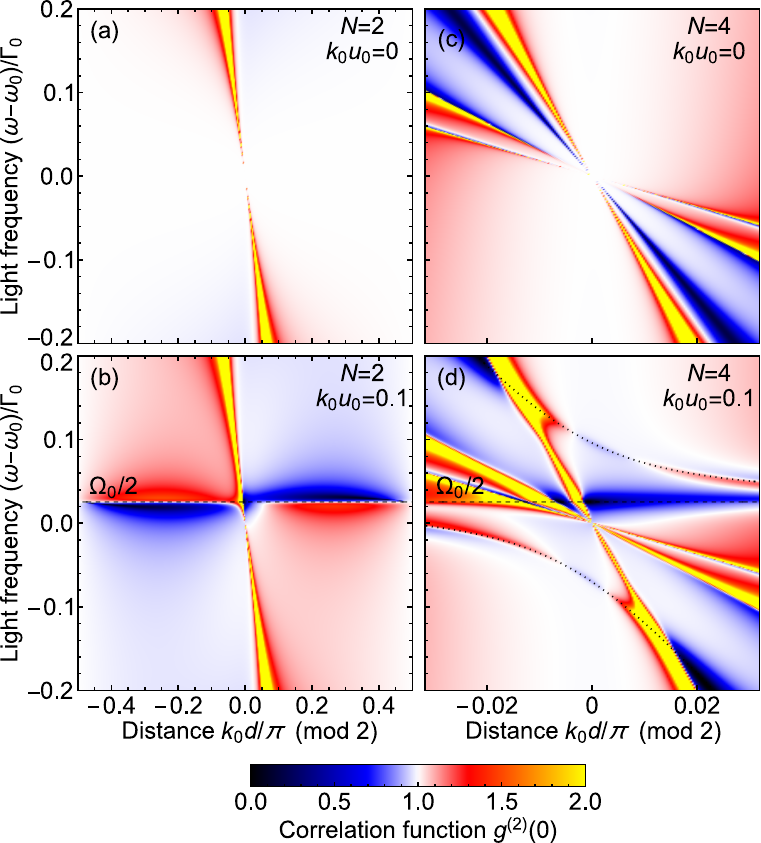}
    \caption{Second-order correlation function of the light reflected from arrays of (a),(b) $N=2$ and (c),(d) $N=4$ qubits, shown as a function of the incident light frequency $\omega$ and the distance between the qubits $d$. Panels (a) and (c) show the case with no interaction with the mechanical motion of the qubits, while panels (b) and (d) include this interaction. Calculations were performed using Eqs.~\eqref{eq:g2M},~\eqref{eq:M0}, and~\eqref{eq:Mv} for $\Omega_0 - 2\omega_0 = 0.05 \Gamma_0$ and other parameters indicated in the figure.}
    \label{fig:g24}
\end{figure}

For a single qubit, the frequency-mixing part of the scattering matrix reads
\begin{align}
&M_{--,++}(\omega_1',\omega_2'; \omega_1, \omega_2) = 2 s(\omega_1) s(\omega_2) [s(\omega_1') + s(\omega_2')] \nonumber\\
&\times \left( 1 + \frac{k_0^2 u_0^2}{2} \frac{\omega_1 + \omega_2 - 2 \omega_0 + 2 \rmi \Gamma_0}{\omega_1 + \omega_2 - \Omega_0 + \rmi k_0^2 u_0^2 \Gamma_0} \right) \,,
\end{align}
where $s(\omega) = 1 / (\omega - \omega_0 + \rmi \Gamma_0)$, which gives the correlation function
\begin{align}\label{eq:g2N1}
&g^{(2)}(\tau) =\\ \nonumber
&\left| 1 - \left( 1 + k_0^2 u_0^2 \frac{\omega - \omega_0 + \rmi \Gamma_0}{2 \omega - \Omega_0 + \rmi k_0^2 u_0^2 \Gamma_0} \right) \e^{-(\rmi \omega_0 + \Gamma_0) |\tau|} \right|^2 \,.
\end{align}
Figure~\ref{fig:g1} shows the dependence of the correlation function at zero delay $g^{(2)}(0)$ on the incident light frequency $\omega$ and the mechanical frequency $\Omega_0$. When the double light frequency is far detuned from the mechanical resonance, $|2\omega - \Omega_0| \gg k_0^2 u_0^2 \Gamma_0, |\Omega_0 - 2 \omega_0|$, the interaction with mechanical motion is suppressed, and the correlation function tends to zero (blue color in Fig.~\ref{fig:g1}), as expected for a single two-level system~\cite{Shen2007}. In contrast, when the two-level qubit couples to the bosonic degree of freedom corresponding to its mechanical motion, bunching occurs (red color in Fig.~\ref{fig:g1}).

Next, we discuss the correlation function at zero delay $g^{(2)}(0)$ for arrays with two and four qubits. To determine the effect of mechanical motion, we plot the correlation function in Fig.~\ref{fig:g24} for cases with and without light-motion interaction. We focus on the region of small $k_0 d$ (mod $2\pi$), where strong signatures of dark states are present. For the two-qubit case, resonance with the single-excited dark states leads to the suppression of the reflection coefficient $t_{-+}$ in the denominator of Eq.~\eqref{eq:g2M}, resulting in strong bunching of the reflected photons, as shown in Fig.~\ref{fig:g24}(a). When mechanical motion is included, an additional resonant feature appears in the $g^{(2)}(0)$ dependence at $\omega = \Omega_0 / 2$, indicated by the horizontal dashed line in Fig.~\ref{fig:g24}(b).

In the four-qubit array, several single- and double-excited states lead to photon bunching and anti-bunching regions, as seen in Fig.~\ref{fig:g24}(c). Including interactions with vibrations introduces a resonance at $\omega = \Omega_0 / 2$, originating from mechanical modes that weakly interact with the light. Additionally, there appears a characteristic avoided crossing pattern, marked by the dotted line in Fig.~\ref{fig:g24}(d). This feature results from resonance with the biphonoritonic modes formed by strongly interacting photon pairs and vibrations, as described at the end of Sec.~\ref{sec:M}.

\section{Summary}\label{sec:Sum}

To summarize, we theoretically explored a broad spectrum of relativistic quantum effects in the waveguide quantum electrodynamics setup, driven by the motion of the qubits  with respect to the waveguide. The dynamical Casimir and Unruh effects, which manifest as light emission and qubit excitation, as well as the radiation recoil are analyzed under conditions of parametric resonance, when the motion frequency of the qubits is close to twice the atomic resonance frequency. Our findings reveal that, depending on the number of qubits and their motion patterns, the emitted photons can be squeezed, have strong directivity, or be directionally entangled. Additionally, the steady state of the moving qubits is found to be entangled, with the potential to undergo quantum dissipative phase transitions. Optomechanical backaction is notably enhanced when the mechanical motion frequency coincides with that of a double-excited subradiant state, leading to the formation of hybrid biphonoriton states---superpositions of a phonon and a pair of photons.

Our study is based on the developed perturbative diagrammatic technique and the rigorous master equation approach. These approaches are complementary, being efficient for weak parametric interaction and an arbitrary number of qubits, or strong interaction and a small number of qubits, respectively. While our focus was on qubits coupled to a one-dimensional waveguide, the methods can be easily extended to setups in free space or inside structured dielectric environments by incorporating appropriate photon Green's functions. Further generalizations could include replacing the two-level qubits with multi-level atoms (e.g, in V, $\Lambda$ or $\Xi$ configuration) or considering rotational motion at a constant angular velocity that meets the parametric resonance condition.

We anticipate that the proposed effects can be experimentally realized by simulating qubit motion in state-of-the-art superconducting transmission lines coupled to transmons with tunable parameters~\cite{Redchenko2022}. Other potential avenues for engineering parametric interactions include using polariton condensates in semiconductors~\cite{Kuznetsov_2023} or oscillating molecules in plasmonic cavities~\cite{Baumberg2018}. The theory developed here is applicable to a whole range of quantum nonlinear systems that are parametrically coupled to a continuum of modes.


\appendix

\section{Diagrammatic technique}\label{app:Dia}

\begin{figure}
    \centering
    \includegraphics[width=0.99\columnwidth]{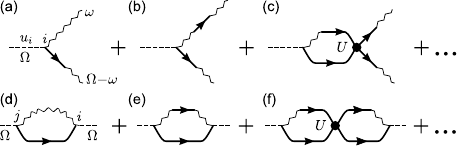}
    \caption{Diagrammatic representation of the contributions to (a)--(c) the two-photon emission amplitude and (d)--(f) the self-energy of the mechanical motion. Straight solid lines represent qubit excitations and are described by the Green's function, Eq.~\eqref{eq:Gx}. Wavy lines represent photons, described by the Green's function $D_{ij}(\omega) = \e^{\rmi k_0 |z_i-z_j|}/(\omega-\omega_k + \rmi 0)$. Dashed lines represent qubit vibrations. Bold dots indicate the fictitious interaction of the qubit excitations, as given in Eq.~\eqref{eq:U}. Other vertices correspond to the interaction Hamiltonians in Eqs.~\eqref{eq:V0}, \eqref{eq:V1}.}
    \label{fig:dia}
\end{figure}

Here, we describe the diagrammatic technique used to perturbatively calculate the DCE emission (Sec.~\ref{sec:DCEperp}) and the DCE recoil effects (Sec.~\ref{sec:Back}). In this technique~\cite{Ke2019,Sheremet2023,Vyatkin2023}, the atomic operators $b_n$ of the two-level system are replaced with corresponding bosonic operators, allowing the application of Wick's theorem. To restore the original level structure, a fictitious nonlinear interaction term is added to the Hamiltonian~\cite{Baranger2013,Poshakinskiy2016,Ke2019},
\begin{equation}\label{eq:U}
    U = \frac{\chi}{2} \sum_{n}b_n^\dagger b_n^\dagger b_n b_n \,.
\end{equation}
In the final equations, the limit $\chi \to \infty$ is taken. In this limit, all states with two or more excitations of one qubit acquire infinite energy and are thus unpopulated.

The diagrams contributing to the process of two-photon generation by an array of moving qubits are shown in Fig.~\ref{fig:dia}(a,b,c). Initially, the parametric term $V_1$, Eq.~\eqref{eq:V1}, creates a photon and a qubit excitation, which can be emitted as a second photon, Fig.~\ref{fig:dia}(a). Alternatively, the former photon can be converted into a second qubit excitation, Fig.~\ref{fig:dia}(b,c). The two excitations can then interact with each other due to the qubit nonlinearity, Eq.~\ref{eq:U}, before finally being converted into a pair of photons, Fig.~\ref{fig:dia}(c). The resulting total two-photon amplitude is given by $\Psi_{\sigma_1,\sigma_2}(\omega_1,\omega_2) =  \psi_{\sigma_1,\sigma_2} (\omega_1,\omega_2) \times 2\pi \delta(\omega_1+\omega_2-\Omega)$, with
\begin{align}\label{eq:psiN}
    &\psi_{\sigma_1,\sigma_2}(\omega_1,\omega_2)=- \omega_0\Gamma_0   
    \Big\{       
    \rmi \Gamma_0\sum_{ijlm}  (u_l-u_m) \phi_{lm} \\  &\times s_i^{(\sigma_1)}(\omega_1)s_j^{(\sigma_2)}(\omega_2)    
 [(\Omega-\mathcal{H}^{(2)})(\Omega-\mathcal{H}^{(2)}-\mathcal{U})^{-1}]_{ij,lm} 
\nonumber   \\ \nonumber
    &+ \sum_{m}u_m [\sigma_1 s_m^{(\sigma_2)}(\omega_2)e^{-ik_0 \sigma_1 z_m}+\sigma_2 s_m^{(\sigma_1)}(\omega_1)e^{-ik_0 \sigma_2 z_m}]  \Big\} .
\end{align}
Here, the last line corresponds to the diagram in Fig.~\ref{fig:dia}(a), while the second and third lines correspond to the sum of diagrams in Fig.~\ref{fig:dia}(b,c). The latter contribution describes the two-photon resonances and is present only if there are at least two qubits with different $u_n$.
In Eq.~\eqref{eq:psiN}, we used the amplitude of the photon emission by the $i$-th qubit
\begin{align}
    &s_i^{(\sigma)}(\omega) = \sum_j G_{ij} (\omega) \e^{-\rmi \sigma k_0 z_j} ,
\end{align}
the Green's function and the effective Hamiltonian of the atomic excitations
\begin{align}\label{eq:Gx}
    &\bm{G}(\omega) = (\omega- \bm{H}_{\rm eff})^{-1} ,\\
    &H^{\rm (eff)}_{ij} = \omega_0 \delta_{ij} - \rmi\Gamma_0 \e^{\rmi k_0 |z_j-z_j|} ,
\end{align}
the Hamiltonian for two independent excitations
\begin{align}
    \mathcal{H}^{(2)} = H_{\rm eff} \otimes 1 + 1 \otimes H_{\rm eff} ,
\end{align}
the matrix describing the fictitious interaction of two excitations, Eq.~\eqref{eq:U},
\begin{align}
    \mathcal{U}_{ij,kl} = \chi \delta_{ij} \delta_{kl} \delta_{ik} \,,
\end{align}
and the phase factor
\begin{align}
    &\phi_{lm} = \text{sign}(z_l-z_m)\,e^{ik_0|z_l-z_m|} . \label{eq:phi}
\end{align}
Note that the sign function appears in Eq.~\eqref{eq:phi} due to the chosen form of $g^{(1)}_{k,n}$, Eq.~\eqref{eq:g1mo}, which is odd in $k$ and corresponds to the qubit motion along the waveguide, considered here. In the case of motion perpendicular to the waveguide, the sign factor in Eq.~\eqref{eq:phi} would be absent. 


\begin{figure}
    \centering
    \includegraphics[width=0.99\columnwidth]{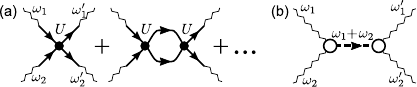}
    \caption{Diagrammatic representation of the two contributions to the two-photon scattering amplitude, which correspond to Eqs.~\eqref{eq:M0} and~\eqref{eq:Mv}, respectively. The thick dashed line represents the dressed Green's function of the vibration $\bm{\mathcal{G}}$ given by Eq.~\eqref{eq:Gm}. Open circles represent the sum of all diagrams in Fig.~\ref{fig:dia}(a)--(c). Other notations are the same as described in the caption of Fig.~\ref{fig:dia}.}
    \label{fig:dia2}
\end{figure}

Now, we calculate the self-energy of the mechanical motion $\Sigma_{ij}(\Omega)$, which we use to obtain the total emission rate $W$ according to Eq.~\eqref{eq:WN}. The contributions to $\Sigma_{ij}(\Omega)$ are shown by the diagrams in Fig.~\ref{fig:dia}(d,e,f) and yield
\begin{align}\label{eq:sigN}
    \Sigma_{ij}(\Omega)=\Gamma_0k_0^2 \Big[   
     \sum_{ml}\Big(\frac{\Gamma_0}{\Omega-\mathcal{H}^{(2)}-\mathcal{U}}\Big)_{il,(jm)}  \phi_{il} \phi_{jm} - \rmi \delta_{ij} \Big],
\end{align}
where the indices in the brackets indicate symmetrization, $A_{(ij)} = A_{ij} + A_{ji}$. In Eq.~\eqref{eq:sigN}, the second term corresponds to the diagram in Fig.~\ref{fig:dia}(d), and the first term corresponds to the sum of the diagrams in Fig.~\ref{fig:dia}(e,f). The expression for the emission rate can be further simplified using the expansion over the double-excited eigenstates of the array,
\begin{align}\label{eq:eigexp}
\left(\frac{1}{\Omega-\mathcal{H}^{(2)}-\mathcal{U}}\right)_{il,(jm)} = 2 \sum_\nu \frac{\psi^{(\nu)}_{il} \psi^{(\nu)}_{jm}}{\Omega - E^{(\nu)}} 
\end{align}
where $\psi^{(\nu)}_{il}$ and $E^{(\nu)}$ are the eigenfunctions and (complex) eigenenergies of the two-excitation problem, 
\begin{align}
\sum_{mj} (\Omega-\mathcal{H}^{(2)}-\mathcal{U})_{il,mj} \psi^{(\nu)}_{mj}= E^{(\nu)} \psi^{(\nu)}_{il} \,,
\end{align}
which are symmetric, i.e., $\psi^{(\nu)}_{il}= \psi^{(\nu)}_{li}$, and should be normalized according to $\sum_{il} \psi^{(\nu)}_{il} \psi^{(\nu')}_{il} = \delta_{\nu,\nu'}$. Substituting the expansion Eq.~\eqref{eq:eigexp} into Eq.~\eqref{eq:sigN}, we obtain Eq.~\eqref{eq:sigNeig}.

Finally, we describe the two-photon scattering in the optomechanical system, where the qubits can move along the waveguide (see Sec.~\ref{sec:g2}). The process is characterized by the scattering matrix
\begin{align}\label{eq:Swwww}
&S_{\sigma_1'\sigma_2' ,\sigma_1\sigma_2}(\omega_1',\omega_2' ; \omega_1, \omega_2) =\\\nonumber
&  (2\pi)^2  t_{\sigma_1' \sigma_1}(\omega_1)t_{\sigma_2' \sigma_2}(\omega_2)\delta(\omega_1'-\omega_1)\delta(\omega_2'-\omega_2)+ (1' \leftrightarrow 2') \\\nonumber
&+2\pi \rmi M_{\sigma_1'\sigma_2' ,\sigma_1\sigma_2}(\omega_1',\omega_2' ; \omega_1, \omega_2) \delta(\omega_1'+\omega_2' -\omega_1-\omega_2) ,
\end{align}
where the quantities without and with a prime correspond to the incident and scattered photons, respectively. The second line in Eq.~\eqref{eq:Swwww} corresponds to the independent transmission or reflection of the two photons, described by the usual coefficients
\begin{align}
t_{\sigma' \sigma}(\omega) = \delta_{\sigma\sigma'} -\rmi \Gamma_0 \sum_{ij} G_{ij}(\omega)\, \e^{\rmi k_0 ( \sigma z_j - \sigma' z_i)} .
\end{align}
The last line in Eq.~\eqref{eq:Swwww} describes the frequency-mixing processes and consists of two contributions $M = M^{(0)} + M^{(v)}$, where
\begin{align}\label{eq:M0}
&M^{(0)}_{\sigma_1'\sigma_2' ,\sigma_1\sigma_2}(\omega_1',\omega_2' ; \omega_1, \omega_2) =- 2\sum_{ij} s_i^{\sigma_1'}(\omega_1')s_i^{\sigma_2'}(\omega_2')   \nonumber \\
&\times [U(\Omega-H^{(2)}-U)^{-1}(\Omega-H^{(2)})]_{ii,jj}
 s_j^{-\sigma_1}(\omega_1)s_j^{-\sigma_2}(\omega_2)
\end{align}
is the vibration-independent part, given by the diagrams in Fig.~\ref{fig:dia2}(a), which arise due to the nonlinearity of the qubits~\cite{Ke2019}, and 
\begin{align}\label{eq:Mv}
&M^{(v)}_{\sigma_1'\sigma_2' ,\sigma_1\sigma_2}(\omega_1',\omega_2' ; \omega_1, \omega_2)  \\ \nonumber
&=   \sum_{ij} \mathcal{G}_{ij}(\omega_1+\omega_2)\,\psi_{\sigma_1'\sigma_2'}^{(i)}(\omega_1',\omega_2') \, \psi^{(j)}_{-\sigma_1,-\sigma_2}(\omega_1,\omega_2)
\end{align}
is the contribution from the processes involving the conversion of a pair of photons into a vibration followed by their reemission, as shown in Fig.~\ref{fig:dia2}(b). Here, the matrix ${\mathcal{G}}_{ij}$ is given by Eq.~\eqref{eq:Gm}, and $\psi^{(i)}$ is given by Eq.~\eqref{eq:psiN}, where one should set $u_l = u_0 \delta_{li}$, i.e., consider the zero-point motion of the $i$-th qubit. Note that the above expression for the scattering matrix~\eqref{eq:Mv} is correct only up to the order $k_0^2u_0^2$, since the initial interaction Hamiltonian in Eqs.~\eqref{eq:V1} and \eqref{eq:g1mo} was limited to the order $k_0u_0$. However, it is still important to retain the higher orders in $k_0u_0$ in Eq.~\eqref{eq:Mv}, as they ensure the unitarity of the scattering matrix.

\section{Derivation of the master equation with Keldysh formalism}\label{app:Keld}

Here, we employ the Keldysh formalism to derive the master equation for an array of qubits that are parametrically coupled to waveguide photons. This approach is more general than the one presented in Sec.~\ref{sec:mast} and can be also used to describe the systems where the couplings are modulated at a set of commensurate or incommensurate frequencies.

Notably, the Keldysh diagram technique~\cite{Keldysh1965}, originally developed for fermions and bosons, cannot be directly applied to two-level atoms~\cite{Arseev2015}. Instead, we utilize the formalism based on the Keldysh action~\cite{Kamenev2011}. For the system under consideration, the full Keldysh action is given by
$S = S_{\rm ph} + S_{\rm qb} + S_{\rm int}$, where
\begin{align}
&S_{\rm ph} =  \sum_{k}  \int  dt \\\nonumber
&\left[ \begin{array}{c}  a_{k,+}(t) \\ a_{k,-}(t) \end{array}\right]^\dag \left[ \begin{array}{cc}  \rmi\partial_t -\omega_k + \rmi\gamma & 0 \\ -2\rmi\gamma &  -\rmi\partial_t +\omega_k + \rmi\gamma \end{array}\right] \left[ \begin{array}{c}  a_{k,+}(t) \\ a_{k,-}(t) \end{array}\right] 
\end{align} 
describes the free electromagnetic field at zero temperature, $S_{\rm qb}$ describes the qubits and depends solely on their variables, while the interaction is given by
 \begin{align}
S_{\rm int} = &\sum_{k,n} \int dt \,g_{n,k}^*(t)  \left\{ [b_{n,+}(t) + b_{n,+}^\dag(t)] a_{k,+}(t) \, \e^{\rmi k z_n} \right.  \nonumber \\
- &\left.  [b_{n,-}(t) + b_{n,-}^\dag(t)] a_{k,-}(t) \, \e^{\rmi k z_n} \right\} + {\rm h.c.} 
 \end{align}
Here, we use the notation where the indices $+$ and $-$ correspond to the forward and backward parts of the Keldysh contour, and a small positive $\gamma \to 0$ is introduced to ensure convergence.

Next, we integrate $\e^{\rmi S}$ over the photonic variables. Due to the quadratic form of the action, the integrals are straightforward to evaluate, yielding $\e^{\rmi (S_{\rm qb}+S_{\rm eff})}$. The effective action correction is given by
\begin{align}\label{eq:SeI}
 S_{\rm eff} =  - &\sum_{k,n,m} \int_{-\infty}^\infty \frac{d\omega}{2\pi}\,  \e^{\rmi k (z_n-z_m)} 
 \\\nonumber
& \left[ \begin{array}{c}  p_{n,k,+}(\omega) \\ -p_{n,k,-}(\omega) \end{array}\right]^\dag \hat D_k(\omega) \left[ \begin{array}{c}  p_{m,k,+}(\omega) \\ -p_{m,k,-}(\omega) \end{array}\right] \,,
\end{align}
where we have switched to the frequency domain and introduced the qubit polarizations
\begin{align}
p_{n,k,\pm}(t) = g_{n,k}(t)  [b_{n,\pm}(t) + b_{n,\pm}^\dag(t)] \,,
\end{align}
the corresponding Fourier spectra $p_{n,k,\pm}(\omega)$, and the photon Green's function
\begin{align}\label{eq:Gph}
\hat D_k(\omega) =  \left[ \begin{array}{cc}  \dfrac1{\omega -\omega_k + \rmi 0} & 0 \\ -2\pi\rmi\delta(\omega-\omega_k) &  -\dfrac1{\omega -\omega_k - \rmi 0} \end{array}\right]  \,.
\end{align}

When evaluating the sum over $k$ in Eq.~\eqref{eq:SeI}, we assume that the dominant contribution comes from $k \approx k_0$, where $\omega_{k_0}$ matches the qubit's resonant frequency, consistent with the rotating-wave approximation. We also decompose $g_{n,k}$ and $p_{n, k}$ into $k$-even and $k$-odd parts:
\begin{align}
&g_{n,\pm k_0} = g_{n, e} \pm g_{n, o} \,,\\
&p_{n, \pm k_0} = d_n \pm q_n \,.
\end{align}
This leads to
\begin{align}\label{eq:SeI2}
 S_{\rm eff} =  -\rmi  \sum_{n,m} \int\limits_0^\infty \frac{d\omega}{2\pi c}
& \left[ \begin{array}{c}  d_{n,+}(\omega) \\ q_{n,+}(\omega) \\ d_{n,-}(\omega) \\ q_{n,-}(\omega) \end{array}\right]^\dag \hspace{-.3em}
\hat T_{nm} \hspace{-.1em}
\left[ \begin{array}{c}  d_{m,+}(\omega) \\ q_{m,+}(\omega) \\ d_{m,-}(\omega) \\ q_{m,-}(\omega) \end{array}\right]
\end{align}
where 
\begin{align}
\hat T_{nm} &= \left[ \begin{array}{cc}
-\rmi H_{nm} & 0 \\ 
J_{nm} & \rmi H_{mn}^*
\end{array}\right] \,,\\ \nonumber
H_{nm} &= -\rmi \,\e^{\rmi k_0|z_n-z_m|} 
 \left[ \begin{array}{cc}
1 & {\rm sign}(z_n-z_m) \\
{\rm sign}(z_n-z_m) & 1
\end{array}\right] \,,\\ \nonumber
J_{nm} &= 2
 \left[ \begin{array}{cc}
\cos k_0(z_n-z_m) & \rmi \sin k_0(z_n-z_m) \\
\rmi \sin k_0(z_n-z_m) & \cos k_0(z_n-z_m)
\end{array}\right] \,,
\end{align}
with $c = d\omega_k/dk$ at $k=k_0$ and ${\rm sign\,}0=0$ assumed. Note that only frequencies $\omega>0$ contribute to the integral in Eq.~\eqref{eq:SeI2}, since the Green's function in Eq.~\eqref{eq:Gph} has poles only for positive $\omega$.

We take the coupling modulation 
\begin{align}
g_{n, e(o)}(t) =  g^{(0)}_{n,e(o)}  + g^{(1)}_{n,e(o)}  \e^{-\rmi \Omega t}  \,,
\end{align}
where $\Omega$ is close to twice the qubit resonant frequency, and take the positive-frequency part of $d_{n,\pm}$, $q_{n,\pm}$. 
{Then, the effective action assumes the form $S_{\rm eff} = \int L_{\rm eff} dt$ where 
\begin{align}\label{eq:Se}
& L_{\rm eff} =  -\rmi  \sum_{n,m} 
\left[ \begin{array}{c}  
p_{n,e,+}(t) \\ 
p_{n,o,+}(t) \\
p_{n,e,-}(t) \\ 
p_{n,o,-}(t) 
\end{array}\right]^\dag  \,
\hat T_{nm} 
\left[ \begin{array}{c}  
p_{n,e,+}(t) \\ 
p_{n,o,+}(t) \\
p_{n,e,-}(t) \\ 
p_{n,o,-}(t) 
\end{array}\right] 
 \,, \\ \nonumber
&p_{n,e(o),\pm}(t)= g_{n,e(o)}^{(0)}b_{n,\pm}(t)+ g_{n,e(o)}^{(1)}b_{n,\pm}^\dag(t) \,\e^{-\rmi\Omega t} .
\end{align}
Since the action is time-local, it corresponds to the Markovian dynamics of the qubit density matrix, governed by $d\rho = d\rho_{\rm qb} + \rmi \mathcal{T_C} L_{\rm eff}(t) \rho(t) dt$, where $d\rho_{\rm qb}$ represents the free evolution of the qubits and the operation $\mathcal{T_C}$  puts all the operators with $+$($-$) index to the left(right) side of $\rho(t)$. This yields the master equation}
\begin{align}\label{eq:KdrhoL}
\frac{d\rho}{dt} = -\rmi (H_{\rm eff} \rho - \rho H_{\rm eff}^\dag) + \mathcal{J}[\rho]
\end{align}
with
\begin{align}
 H_{\rm eff}&=H_{\rm qb}-\frac{\rmi}c \sum_{n,m} \Big[ p_{n,e}^\dag p_{m,e} +p_{n,o}^\dag p_{m,o}  \\[-2mm] \nonumber
& +\left(p_{n,e}^\dag p_{m,o} +p_{n,o}^\dag p_{m,e}\right) {\rm sign}(z_n-z_m) \Big] \e^{\rmi k_0|z_n-z_m|} \,,\\
\mathcal{J}[\rho]& = \frac2c \sum_{n,m} \Big[  \left(p_{m,e} \rho p_{n,e}^\dag + p_{m,o} \rho p_{n,o}^\dag\right) \cos k_0(z_n-z_m) \nonumber \\[-2mm] \label{eq:KJ}
 +&\rmi  \left(p_{m,e} \rho p_{n,o}^\dag  + p_{m,o}\rho p_{n,e}^\dag \right)  \sin k_0(z_n-z_m)  \Big] \,,
\end{align}
where $H_{\rm qb}$ is the Hamiltonian of the free qubits. Finally, Eqs.~\eqref{eq:KdrhoL}-\eqref{eq:KJ} can be rewritten in the form of Eqs.~\eqref{eq:drhoL}-\eqref{eq:J}.

\section{Emission characteristics}\label{app:Emi}

Here, we describe the approach used to calculate the properties of light emitted by an array of moving qubits in the non-perturbative regime. We begin with the input-output relation for the photon operators~\cite{Gardiner1985,Carmichael1987}:
\begin{align}\label{eq:IO}
&a_{\sigma}^{\rm out} (t)= a_{\sigma}^{\rm in} (t) - \rmi B_{\sigma} (t)\,,\\ 
&B_{\sigma} =\sum_n B_{n, \sigma k_0} \e^{-\sigma \rmi k_0 z_n}\,,
\end{align}
where $a^{\rm in,out}_{\sigma} (t) = \sum_{\sigma k > 0} a_k^{\rm in,out} e^{-\rmi\omega_k t}$ is the field operator for photons moving in the direction $\sigma = \pm$. The emission properties are then expressed via the correlation functions of the $B_{\sigma}$ operators, which can be evaluated using the quantum regression theorem, based on the steady-state density matrix of the qubits $\rho_\infty$.  

In particular, the spectrum of the emission in the direction $\sigma = \pm$ is given by
\begin{align}
I_{\sigma}(\nu) =  
2 {\,\rm Re\,} {\rm Tr} \{ B_{\sigma}^\dag (\rmi\nu - \mathcal{L})^{-1}[B_{\sigma} \rho_\infty] \} \,,
\end{align}
where $\nu  = \omega-\Omega/2$ is the frequency in the rotating frame and $\mathcal{L}$ is the Liouvillian, Eq.~\eqref{eq:drhoL}. The (unnormalized) directional density matrix of the emitted photons in Eq.~\eqref{eq:rho1p} is given by
\begin{align}
J_{\sigma\sigma'}^{(1)} = {\rm Tr} (B_{\sigma'}^\dag B_{\sigma} \rho_\infty) 
\end{align}
and the total emission rate in the direction $\sigma$ is $W_{\sigma} = J_{\sigma\sigma}^{(1)}$. The second-order correlation function  of the light is calculated as
\begin{align}
g_{\sigma}^{(2)}(t) =\frac{1}{W_{\sigma}^2}\, {\rm Tr} \{ B_{\sigma}\, \e^{\mathcal{L}t} [B_{\sigma} \rho_\infty B_{\sigma}^\dag] B_{\sigma}^\dag\} \,.
\end{align}
The wave function of a pair of emitted photons, Eq.~\eqref{eq:psi2u}, reads
\begin{align}\label{eq:psi2B}
&\psi_{\sigma\sigma'}(\tau) = \theta(\tau)\tilde\psi_{\sigma\sigma'}(\tau)+\theta(-\tau)\tilde\psi_{\sigma'\sigma}(-\tau) \,,\\ \nonumber
&\tilde\psi_{\sigma\sigma'}(\tau) =  - {\rm Tr} \{ B_{\sigma} \e^{\mathcal{L}\tau}[B_{\sigma'} \rho_\infty] \} \,,
\end{align}
which can be further used to calculate the two-photon directional density matrix and concurrence. 

To describe the field squeezing, we introduce the field quadratures $X_{\sigma,\theta}(t) =  \e^{-\rmi\theta} a_{\sigma}^{\rm out}(t)  +   \e^{\rmi\theta} a_{\sigma}^{{\rm out}\,\dag} (t)$ and calculate the symmetrized correlation function
\begin{align}
&K_{\sigma\theta,\sigma'\theta}(t) \equiv \langle \{ X_{\sigma,\theta}(t),X_{\sigma',\theta}(0) \} \rangle 
=\delta(t) \\
  &+ 2\, {\rm Re}\,\langle a_{\sigma}^{{\rm out}\,\dag}(t) a_{\sigma'}^{\rm out}(0)\rangle \nonumber
   + 2\, {\rm Re}\, \langle \mathcal{T} a_{\sigma}^{\rm out}(t) a_{\sigma'}^{\rm out}(0)\rangle\, \e^{-2\rmi \theta} 
 \end{align}
 where $\{A,B\} = (AB+BA)/2$, $\mathcal{T}$ denotes time ordering, and the canonical commutation relations for the output field were used. Substituting Eq.~\eqref{eq:IO}, we obtain  
 \begin{align}
K_{\sigma\theta,\sigma'\theta}(t) =
\delta(t) 
 +& 2\, {\rm Re}\,\langle B_{\sigma}^{\dag}(t) B_{\sigma'}(0)\rangle  \\ \nonumber
   - &2\, {\rm Re}\, \langle \mathcal{T} B_{\sigma}(t) B_{\sigma'}(0)\rangle\, \e^{-2\rmi \theta} 
 \end{align}
 where we have used $\langle \mathcal{T} B_{\sigma}(t) a_{\sigma'}^{\rm in} (0) \rangle = 0$, which is trivial for $t>0$, and follows from the causality commutation relation $ [B_{\sigma}(t), a_{\sigma'}^{\rm in} (0)]=0$  for $t<0$~\cite{Gardiner1985}. Finally, the field quadrature noise is given by the Fourier spectrum of the correlation function with $\sigma=\sigma'$,
\begin{align}
K_{\sigma\theta,\sigma\theta}(\nu) = 1 + 2\, {\rm Re}[\e^{-2\rmi \theta} \psi_{\sigma\sigma}(\nu) ] + I_{\sigma}(\nu)+I_{\sigma}( - \nu)
\end{align}
where $\psi_{\sigma}(\nu)$ is the Fourier spectrum of the two-photon wave function $\psi_{\sigma\sigma}(t)$, Eq.~\eqref{eq:psi2B}, given by
\begin{align}
&\psi_{\sigma\sigma}(\nu) = \tilde\psi_{\sigma\sigma}(\nu)+\tilde\psi_{\sigma\sigma}(-\nu)\,, \\ \nonumber
&\tilde\psi_{\sigma\sigma}(\nu)  = 
 {\rm Tr} \{ B_{\sigma} (\rmi\nu + \mathcal{L})^{-1}[B_{\sigma} \rho_\infty] \} \,.
\end{align}
The squeezing parameter for the central frequency $\nu =0$ ($\omega= \Omega/2$) is then found to be
\begin{align}\label{eq:K0}
\xi _{\sigma} = \min_{\theta} K_{\sigma\theta,\sigma\theta}(0) = 
1 - 2 |\psi_{\sigma\sigma}(0)| + 2 I_{\sigma}(0) \,.
\end{align}
One can also consider squeezing of the modes with a certain spatial parity, $ X_{e(o)} = (X_{+} \pm X_{-})/\sqrt2$. The corresponding quadrature correlation functions are given by
\begin{align}
&K_{e\theta,e\theta} = \frac12 \sum_{\sigma\sigma'} K_{\sigma\theta,\sigma'\theta} ,\:\:
&K_{o\theta,o\theta} = \frac12 \sum_{\sigma\sigma'} \sigma\sigma' K_{\sigma\theta,\sigma'\theta} .
\end{align}

 \section{Single qubit moving along the waveguide: non-perturbative approach}\label{app:1mov}

Here, we employ the master-equation approach, described in Sec.~\ref{sec:mast}, to characterize the state and calculate the emission of a single qubit moving along a waveguide beyond the perturbative regime discussed in Sec.~\ref{sec:Low1}.

We assume the coupling parameters are given by $g_{1,\pm k_0} = g_0$ and $g_{1, \pm k_0}^{(1)} = \pm v g_0$, where $v$ is the dimensionless motion amplitude, which is considered real without loss of generality. The dynamics of the density matrix, parametrized by the spin vector $\bm{s}$ as in Sec.~\ref{sec:Mod1}, is governed by
\begin{align}\label{eq:ds1m}
\frac{ds_x}{dt} &= -\Gamma_0 (1+v^2) s_x + \Delta\, s_y \nonumber \,,\\
\frac{ds_y}{dt} &= -\Gamma_0 (1+v^2) s_y - \Delta\, s_x \,,\\
\frac{ds_z}{dt} &= -2\Gamma_0(1+v^2) s_z  - \Gamma_0(1-v^2) \,,\nonumber
\end{align}
where $\Gamma_0 = g_0^2/c$ and $\Delta = \Omega/2 - \omega_0$. While the dynamics of the $s_z$ component is described by the same equation as in the case of motion perpendicular to the waveguide, the relaxation rates of $s_x$ and $s_y$ components differ, cf. Eqs.~\eqref{eq:ds1m} and~\eqref{eq:ds1}. Consequently, the steady state of the qubit, its total emission rate, and the $g^{(2)}$ correlation function, which are determined by $s_z$, match for both cases and are determined by Eqs.~\eqref{eq:nu1},~\eqref{eq:Wu1}, and~\eqref{eq:g2u1}.
However, the emission spectrum, which is determined by $s_x$ and $s_y$ components, is slightly different for the case of motion along the waveguide:
\begin{align}\label{eq:Iu1m}
&I_\sigma (\omega) = 2v^2 \Gamma_0^2
\left[ \frac{1}{(\omega-\omega_0)^2 + \Gamma_v^2 } +  \frac{1}{(\Omega-\omega-\omega_0)^2 + \Gamma_v^2} \right] \,,
\end{align}
cf. Eq.~\eqref{eq:Iu1}. Equation~\eqref{eq:Iu1m} is the extension of the corresponding perturbative result, Eq.~\eqref{eq:I1}, where $k_0 |u_1| = v $ should be set. We find that the only difference from the latter is the replacement $\Gamma_0 \to \Gamma_v \equiv \Gamma_0 (1+v^2)$ in the denominators.

More importantly, we can calculate the full dependence of the squeezing parameter on the motion amplitude $v$,
\begin{align}\label{eq:Xi1um}
\xi_{\sigma} = 1 -  \frac{4v\Big[\sqrt{\Gamma_v^2-\Delta^2 \big(\frac{1-v^2}{1+v^2}\big)^2}-2v\Big]}{\Gamma_v^2+\Delta^2}
\end{align}
for the electric field at the central frequency emitted in the direction $\sigma =\pm$. 
As expected, in the limit $v \ll 1$, Eq.~\eqref{eq:Xi1um} reproduces the series in the perturbative Eq.~\eqref{eq:Xi1ser}. The minimum value of the squeezing parameter, $\xi_{\sigma} = 1/2$, is achieved at $v = 2 \pm \sqrt{3}$. The electric field emitted by the qubit moving along the waveguide is odd with respect to the qubit's central position. The squeezing of this odd mode, $\xi_{o}$, is given by the same Eq.~\eqref{eq:Xi1um}, with the second term on the right-hand side doubled. The dependence of $\xi_{\sigma}$ and $\xi_o$ on the modulation amplitude is shown by the red solid and dashed lines in Fig.~\ref{fig:Xi1}. Note that in the case of the qubit motion along the waveguide, considered here, the squeezing is stronger than in the case of motion perpendicular to the waveguide, cf. the red and blue lines in Fig.~\ref{fig:Xi1}.

%

\end{document}